\documentclass[%
 reprint,
groupedaddress,
 amsmath,amssymb,
 aps,
]{revtex4-2}

\usepackage{graphicx}

 \usepackage{ulem}

\usepackage{dcolumn}
\usepackage{bm}

\usepackage{multirow}

\usepackage{bbding}

\usepackage{braket}

\usepackage{color}

\newcommand\sref[1]{Sec.~\ref{#1}}
\newcommand\eref[1]{Eq.~(\ref{#1})}
\newcommand\fref[1]{Fig.~\ref{#1}}
\newcommand\tref[1]{Tab.~\ref{#1}}

\usepackage{xcolor}
\usepackage{hyperref}

\hypersetup{
    colorlinks,
    linkcolor={blue!80!black},
    citecolor={blue!80!black},
    urlcolor={blue!80!black}
}

\begin{document}

\preprint{APS/123-QED}

\title{
Using the inductive-energy participation ratio to characterize a superconducting quantum chip}

\author{Ke-Hui Yu}
\author{Xiao-Yang Jiao}
\author{Li-Jing Jin}
\email{jinlijing2049@outlook.com}

\affiliation{
 Institute for Quantum Computing, Baidu Research, Beijing 100193, China}

\date{\today}

\begin{abstract}
We developed an inductive-energy participation ratio (IEPR) method and a streamlined procedure for simulating and verifying superconducting quantum chips. These advancements are increasingly vital in the context of large-scale, fault-tolerant quantum computing. Our approach efficiently extracts the key linear and nonlinear characteristic parameters, as well as the Hamiltonian of a quantum chip layout. In theory, the IEPR method provides insights into the relationship between energy distribution and representation transformation. We demonstrate its practicality by applying it to quantum chip layouts, efficiently obtaining crucial characteristic parameters in both bare and normal representations--an endeavor that challenges existing methods. Our work holds the promise of significant enhancements in simulation and verification techniques and represents a pivotal step towards quantum electronic design automation.
\end{abstract}

\maketitle


\section{Introduction}

Benefiting from the rapid advance of micronano and manipulation techniques, 
superconducting quantum chips have emerged as one of the most promising platforms for building quantum computers \cite{arute2019quantum,gong2021quantum,wu2021strong,chow2021ibm,google2023suppressing,alt2022potentials}. 
Despite significant strides \cite{rigetti2012superconducting,nguyen2019high,place2021new,wang2022towards,mcardle2019variational,google2020hartree,eddins2022doubling,mi2022time,zhang2022digital,jafferis2022traversable,ikeda2023first}, challenges persist on the path to realizing large-scale, high-performance quantum chips. In addition to advanced fabrication technology, quantum electronic design automation (QEDA) is a vital domain but remains underestimated.
Analogous to electronic design automation (EDA) in the conventional integrated circuits industry, QEDA plays a pivotal role in advancing large-scale, fault-tolerant quantum computers. It aims to streamline and enhance quantum chip design and verification processes through dedicated tools and methodologies.
Central to these technologies is simulation and verification, pivotal for accurately capturing chip characteristics and ensuring adherence to the design specifications. This process substantially reduces the costs of trial-and-error iterations in chip fabrication.
\begin{figure} [t]
    \centering
    \includegraphics[width=\linewidth]{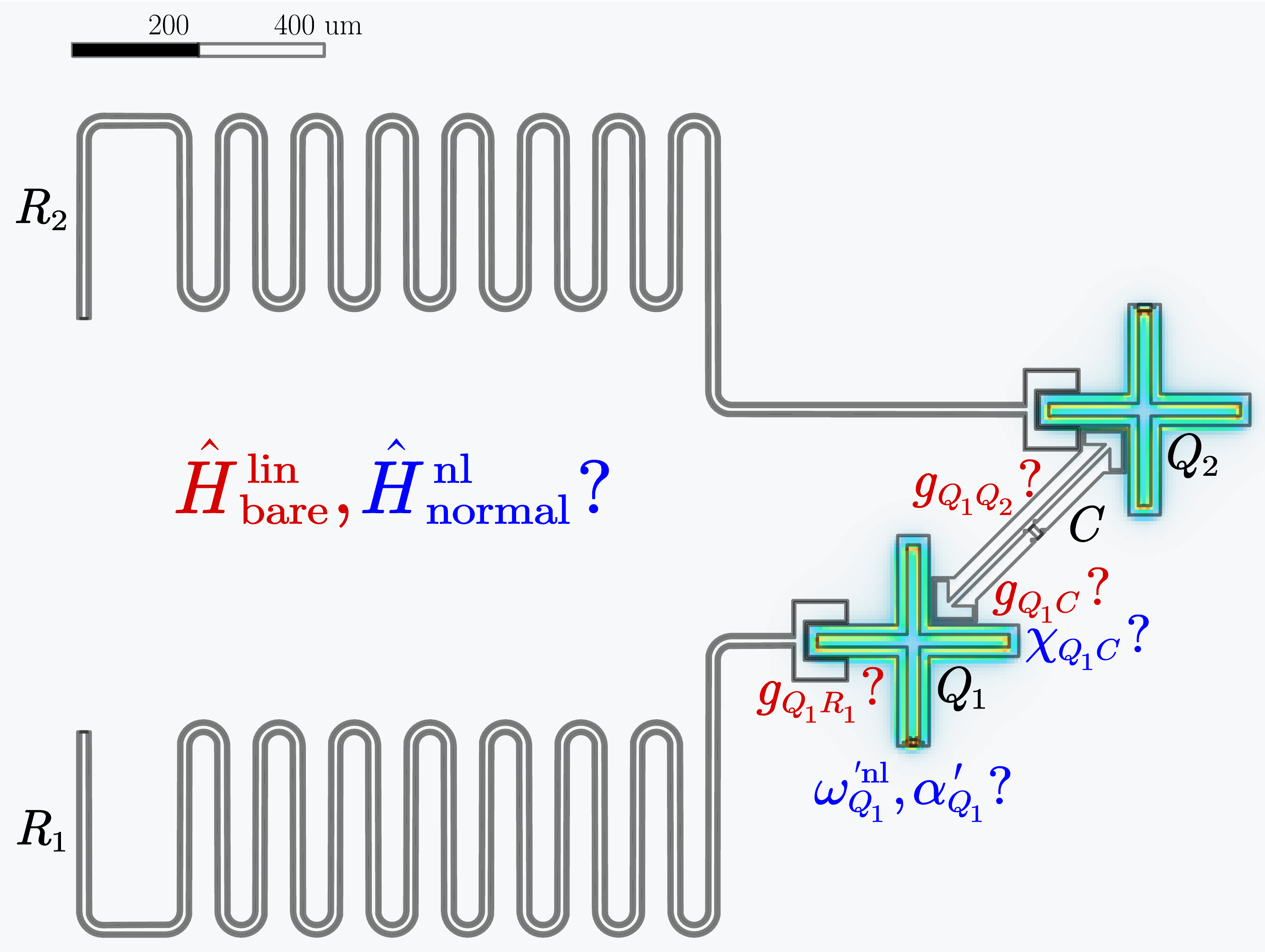}
    \caption{A prototypical quantum chip layout with superconducting coupler architecture. It consists of two qubits ($Q_1,Q_2$), one coupler ($C$), and two resonators ($R_1, R_2$). The electric field distribution can be evaluated via classical electromagnetic simulation. The key question addressed is how to work out efficiently the characteristic parameters of interest both in bare and normal mode, including the linear coupling strength and nonlinear parameters, to construct the interested Hamiltonian, either $\hat H_{\rm bare}^{\rm lin}$ or $\hat H_{\rm normal}^{\rm nl}$.  }
    \label{fig:layout}
\end{figure}

We commence by introducing a prototypical quantum chip layout based on the well-established and widely utilized coupler architecture \cite{arute2019quantum,google2020hartree,google2023suppressing,wu2021strong}, as shown in Fig.~\ref{fig:layout}. It consists of two qubits ($Q_1$, $Q_2$), one coupler ($C$), and two resonators ($R_1$, $R_2$). The nontrivial challenge tackled in this study is to extract the algebraic characteristic parameters (especially from the experimental interest) and the system's Hamiltonian in both bare- and normal-mode representation from the geometric layout.
In the bare-mode representation, aside from the bare frequency of each element, obtaining the linear couplings between different elements is also essential. 
Given the broad spectrum of coupling within quantum chip layouts, spanning from (near) resonant coupling (e.g., $Q_1$-$Q_2$ effective coupling) to dispersive coupling (e.g., $Q_i$-$R_i$ coupling, $i=1,2$, and $Q_i$-$C$ coupling)\cite{regime}, achieving a comprehensive characterization presents a significant challenge. 
Furthermore, due to the nonlinear effect of Josephson junctions, resolving the anharmonicity of relevant elements becomes imperative.
Additionally, in the normal-mode representation (i.e., the laboratory frame), renormalized frequencies, self-Kerr and cross-Kerr parameters assume significance, particularly within the experimental context.
Once obtaining the characterized parameters of interest, the layout's Hamiltonian can be directly derived for both bare- and normal-mode representation. Notably, the bare-mode Hamiltonian encompasses more comprehensive information than its normal-mode counterpart \cite{bare}. 
Yet, obtaining the bare Hamiltonian is regarded as ``challenging and requiring iterative interplay between experimentation and theory" \cite{nigg2012black}.
\par To overcome these challenges, we developed a simple yet nontrivial method termed the ``inductive-energy participation ratio (IEPR)". Compared to the existing techniques, IEPR is applied to figure out all the characterized parameters of interest both in bare- and normal-mode representation and construct the corresponding Hamiltonian of superconducting quantum chip layouts.
Theoretically, we establish a fundamental connection between IEPR and unitary transformation, effectively bridging the gap between the bare mode and normal mode Hamiltonians. We detail the IEPR approach to solving  different types of characteristic parameters in both bare and normal modes. Building upon the IEPR method, we further provide an operational procedure guiding the acquirement of these parameters and the Hamiltonian extracted only from the  quantum chip layout. This methodology empowers us to efficiently model the quantum chip layout's Hamiltonian, offering researchers the means to assess and optimize quantum chip designs before entering the fabrication stage.
The subsequent structure of this paper is as follows: In Sec.~\ref{sec:IEPR}, we present the inductive Energy Participation Ratio method, encompassing both linear and nonlinear regimes. Additionally, we propose a standardized simulation and verification procedure for superconducting quantum chips using the IEPR method in Sec.~\ref{sec:procedure}. Leveraging these methods and procedures, we investigate the linear coupling characteristics, self-Kerr and cross-Kerr parameters, as well as other vital aspects in Sec.~\ref{sec:applications} to demonstrate the method's efficacy and advantages. Finally, we provide a summary of our findings in Sec.~\ref{sec:summary} and include technical details in the Appendices. 

\section{inductive-energy participation ratio method}
\label{sec:IEPR}
The IEPR is defined as follows, offering a quantifiable means to extract characteristic parameters and the Hamiltonian specific to a given superconducting quantum chip layout through classical electromagnetic (EM) simulation.
\begin{equation}
    \label{eq:def_IEPR}
    r_{mn} = \frac{\mathcal E_{mn}^{I}}{\mathcal E_m^{I}},
\end{equation}
where the IEPR $r_{mn}$ embodies the ratio of the average inductive energy $\mathcal E_{mn}^{I}$ distributed across the $n$th element to average inductive energy $\mathcal E_m^{I}$ stored in the full quantum chip when the normal mode $m$ is excited. 
Unlike the EPR method \cite{minev2021energy} which primarily concentrates on capturing inductive energy localized within Josephson junctions, our research reveals that the inductive energy spanning the entire element significantly contributes to obtaining the transformation matrix between the bare and normal Hamiltonian.
Both of these inductive energies can be extracted from classical EM simulation, from which the EM field distribution is evaluated (see Fig.~\ref{fig:layout}). 
Therefore, the main challenge we address turns to leveraging the IEPR to obtain the transformation matrix and further generate the characteristic parameters of interest.

\subsection{The linear regime}
Within the realm of superconducting quantum chips integrating multiple qubits and coplanar resonators, a prevalent model is a capacitively coupled circuit. This investigation initially centers on the linear regime, wherein the determination of coupling strengths among distinct elements takes precedence. Subsequently, we delve into the nonlinear effects in the subsequent subsection.
Considering the frequently used capacitive couplings, the many-body Hamiltonian describing quantum chips in the linear regime is defined as 
\cite{paik2011observation,krantz2019quantum,blais2020quantum,blais2021circuit}
\begin{equation}
    \label{eq:ham_cir}
    \hat{H}_{\rm bare}^{\rm lin} = \sum_m \frac{\hat{Q}_m^2}{2C_m} + \frac{ \hat{\Phi}_m^2}{2L_m} + \frac{1}{2}\sum_{m\neq n}\frac{C_{mn}}{C_m C_n}  \hat{Q}_m \hat{Q}_n,
\end{equation}
where $\hat{Q}_m$ signifies the charge on the $m$th capacitor with capacitance $C_m$, 
while $\hat{\Phi}_m$ pertains to the flux threading the $m$th element's Josephson junctions or inductor with effective inductance $L_m$. Additionally, $C_{mn}$ denotes the mutual capacitance between elements $m$ and $n$. Note that the charge and flux variables are observables having the relation ($\hbar=1$ for simplicity) $[\hat{\Phi}, \hat{Q}]=i$ \cite{blais2021circuit}.
To facilitate further analysis, we introduce an essential technique through the operator replacements: $\hat{\tilde{\Phi}}_m=\hat{\Phi}_m/\sqrt{L_m}$ and $\hat{\tilde{Q}}_m=\hat{Q}_m/(\omega_m \sqrt{C_m})$, with $\omega_m = 1/\sqrt{L_mC_m}$ denoting the bare mode frequency of the $m$th element. Consequently, Hamiltonian \eqref{eq:ham_cir} can be reformulated as:
\begin{equation}
\label{eq:ham_cir_tilde}
 \hat{H}_{\rm bare}^{\rm lin} = \sum_m ({\hat{\tilde{\Phi}}_m^2}/{2} + {\omega_m^2\hat{\tilde{Q}}_m^2}/{2} ) + \sum_{m \neq n} g_{mn}\sqrt{\omega_m \omega_n} \hat{\tilde{Q}}_m \hat{\tilde{Q}}_n,    
\end{equation}
where $g_{mn} = C_{mn}\sqrt{\omega_m \omega_n}/\left(2\sqrt{C_m C_n}\right)$ denoting the coupling strength. Subsequently, we diagonalize the Hamiltonian using a unitary transformation $U$ as $\hat{H}_{\rm normal}^{\rm lin} = U\hat{H}_{\rm{bare}}^{\rm lin} U^\dagger= \sum_m \hat{\tilde{\Phi}}_m'^2/2 + \omega_m'^2 \hat{\tilde{Q}}_m'^2/2$, where $\omega_m'$ signifies the normal mode frequency. The vector space between bare-mode representation $\bm{\Lambda} = (\hat{\tilde{\Phi}}_1, \cdots, \hat{\tilde{\Phi}}_m,\hat{\tilde{Q}}_1, \cdots,\hat{\tilde{Q}}_m)^T$ and normal-mode representation $\bm{\Lambda}'= (\hat{\tilde{\Phi}}'_1, \cdots, \hat{\tilde{\Phi}}'_m,\hat{\tilde{Q}}'_1, \cdots,\hat{\tilde{Q}}'_m)^T$ is also bridged by the unitary transformation, i.e., $\bm{\Lambda}' = U \bm{\Lambda}$. To preserve the commutation relationship $[\hat{\tilde{\Phi}}_m',\hat{\tilde{Q}}_m']=i$ after the transformation, the $U$ matrix assumes a block-diagonal form $\mathcal{U} \oplus \mathcal{U}$, where $\mathcal{U}$ also stands as a unitary matrix with entry $u_{mn}$ (See Apppendix~\ref{sec:relation_IEPR} for more details).
In practical quantum chip layout simulation, although the normal-mode information can be extracted conveniently through classical EM simulation, capturing the bare-mode information (e.g., the coupling strength between different elements) remains essential but nontrivial. Since the normal-mode and bare-mode representations are connected through $U$, the critical question turns to how to figure out $U$ from the normal-mode informations. Herein, we will discern that IEPR holds the key.
\par Utilizing the IEPR definition provided in Eq.~\eqref{eq:def_IEPR} and employing both the bare Hamiltonian $\hat{H}_{\rm bare}^{\rm lin}$ and the normal Hamiltonian $\hat{H}_{\rm normal}^{\rm lin}$, a succinct relationship between IEPR $r_{mn}$ and the unitary transformation $u_{mn}$ emerges. In particular, we derive the compact expression $r_{mn} = {\langle \Psi_m'  | \hat{\tilde{\Phi}}_n^2/2 | \Psi_m'  \rangle} / \sum_k { \langle \Psi_m'  |   \hat{\tilde{\Phi}}_k'^2/2 | \Psi_m' \rangle }$, wherein $| \Psi_m' \rangle$ signifies the eigenstate corresponding to normal mode $m$.
The connection between $\hat{\tilde{\Phi}}_n$ and $\hat{\tilde{\Phi}}_k^\prime$ is governed by the previously mentioned relation $\bm{\Lambda}'=U \bm{\Lambda}$, leading to the equation $\hat{\tilde{\Phi}}_n = \sum_{k} u_{kn} \hat{\tilde{\Phi}}_k'$.
Here,  the subscript $k$ denotes the index of different modes.
Upon substitution, this yields the succinct outcome (see Appendix \ref{sec:relation_IEPR} for more details)
\begin{equation} 
\label{relation:IEPR_U}
r_{mn} = u_{mn}^2.
\end{equation}
To determine $u_{mn}$ from \eref{relation:IEPR_U}, an additional sign matrix $s_{mn}$ is required, which can similarly be acquired through EM simulations. Consequently, the matrix $U$ is expressed as
\begin{equation}
    \label{eq:IEPR_umatrix}
    U = \begin{pmatrix}
        s_{11}\sqrt{r_{11}} &s_{12}\sqrt{r_{12}} &\cdots &s_{1n}\sqrt{r_{1n}}\\
        s_{21}\sqrt{r_{21}} &s_{22}\sqrt{r_{22}} &\cdots &s_{2n}\sqrt{r_{2n}}\\
        \vdots &\vdots &\ddots &\vdots \\
        s_{n1}\sqrt{r_{n1}} &s_{n2}\sqrt{r_{n2}} &\cdots &s_{nn}\sqrt{r_{nn}}
    \end{pmatrix}^{\oplus 2}.
\end{equation}
This matrix reveals the profound relationship between IEPR and representation transformation. The unitary property of matrix $\mathcal{U}$ naturally leads to the orthonormality of IEPR, expressed as $\sum_k s_{nk} s_{mk}\sqrt{r_{nk}r_{mk}}=\sum_k s_{kn} s_{km}\sqrt{r_{kn}r_{km}}=\delta_{mn}$.
Leveraging classical EM simulation, the normal-mode frequencies $\omega_m'$, IEPR $r_{mn}$ and the sign matrix $s_{mn}$ can be calculated, which will be discussed in detail in Sec.~\ref{sec:procedure}. With these parameters, the unitary transformation $U$ can be constructed by \eref{eq:IEPR_umatrix}. As a further step, the bare Hamiltonian is formulated by applying $ \hat{H}_{\rm bare}^{\rm lin} = U^\dagger \hat{H}_{\rm normal}^{\rm lin} U$. Notably, the bare-mode frequency is evaluated as
\begin{align}
    \label{eq:baref}
    \omega_m = \sqrt{\sum_k r_{km} \omega_k'^2},
\end{align}
while the coupling strength is given by
\begin{align}
    \label{eq:coup}
    g_{mn} = \frac{\sum_k s_{km} s_{kn}\sqrt{r_{km} r_{kn}}  \omega_k'^2}{2\sqrt{\omega_m \omega_n}}.
\end{align}

Although the above discussions are based on the capacitive coupling Hamiltonian \eqref{eq:ham_cir}, it is noteworthy that any consideration that can be made for the phase variable can in principle also be made for the charge variable. Therefore, the method can also be extended to the scenarios with inductive couplings if necessary. Most of the techniques would be the same as capacitive couplings, and the main difference is that one has to calculate the capacitive energy participation ratio (CEPR) instead of IEPR. The relevant derivations and discussions are present in Appendix \ref{Appendix:cEPR}.

\subsection{The nonlinear effect}
Given that classical electromagnetic simulation inherently captures the linear electromagnetic attributes of quantum chip layouts, for simplicity, our attention is directed toward the linear characteristics of the quantum chip. However, it is worth noting that the essential nonlinear effects can also be deduced from the linear outcomes. In addition to $\hat{H}_{\rm bare}^{\rm lin}$, we incorporate the nonlinear influence arising from Josephson junctions. Ultimately, it will become evident that pivotal parameters, including renormalized frequencies, self-Kerr and cross-Kerr parameters which commonly hold relevance to real experimental data, can be calculated straightforwardly.
Considering the nonlinear effect of Josephson junctions, the qubit's Hamiltonian is expressed as $\hat H_{q}^{\rm nl} = 4E_C \hat{n}^2 - E_J \cos\hat{\phi}$   \cite{krantz2019quantum,blais2021circuit}
with $E_C$ representing the charging energy and $E_J$ denoting the Josephson energy. The phase operator $\hat{\phi} = 2\pi \hat{\Phi}/\Phi_0$ [$\Phi_0=h/(2e)$ is the flux quantum], and the Cooper-pair number operator $\hat{n} = \hat{Q}/(2e)$ adhere to the commutation relation $[\hat{\phi}, \hat{n}]=i$. In the regime of interest, i.e., $E_C \ll E_J$, we truncate the Josephson energy term to the fourth order while omitting the constant terms, resulting in
\begin{align}
 \label{eq:ham_quantum}
    \hat H_q^{\rm nl} \simeq 4E_C \hat{n}^2 + \frac{E_J}{2}  \hat{\phi}^2  -  \frac{E_J}{24} \hat{\phi}^4.
\end{align}
While the first two terms of the aforementioned equation describe the linear behaviors of the system, the third term introduces the nonlinear effect. In classical electromagnetic simulations, the linear part of Josephson energy term is commonly represented using a lumped inductor with an inductance $L_J$.
The equivalent relation between the Josephson energy and the inductance is governed by $E_J =(\Phi_0/2\pi)^2/L_J$.  
In the absence of the nonlinear effect, the linear qubit frequency is evaluated as $\omega_q=\sqrt{8E_C E_J}$, a value that can be efficiently derived using IEPR from classical EM simulations.
Employing a second quantization $\hat{{\phi}}= (2E_C/E_J)^{1/4} \left( \hat{a}^\dagger +  \hat{a} \right)$ and $\hat{n}=i(E_J/32E_C)^{1/4}\left( \hat{a}^\dagger -  \hat{a} \right)$ to Eq.~\eqref{eq:ham_quantum}, the qubit Hamiltonian is  transformed into the following form:
$\hat H_q^{\rm nl} \simeq \omega_q^{\rm nl} \hat{a}^\dagger \hat{a} + \frac{\alpha_q}{2} \hat{a}^{\dagger 2} \hat{a}^2$, 
where the qubit frequency $\omega_q^{\rm nl} = \omega_q + \alpha_q$ is adjusted by the nonlinear effect, while the qubit anharmonicity $\alpha_q = -E_C$.
Utilizing the aforementioned relationships, we can deduce the expression for the qubit anharmonicity as 
$\alpha_q = - {\omega_q^2 L_J}/{[2(\Phi_0 /\pi)^2]} = - { L_J \sum_k (r_{kq} \omega_k'^2)}/{[2(\Phi_0 / \pi)^2]} $.
Incorporating the nonlinear effects, the bare Hamiltonian describing the many-body system takes the form:
\begin{align}
    \label{eq:many}
    \begin{split}
         \hat H_{\rm bare}^{\rm nl} &= \sum_m \omega_m^{\rm nl} a_m^\dagger a_m + \frac{\alpha_m}{2} a_m^{\dagger 2} a_m^2 \\
         &-\sum_{m\neq n} \frac{g_{mn}}{2} \left(\hat{a}_m^\dagger - \hat{a}_m\right)\left(\hat{a}_n^\dagger - \hat{a}_n\right).
    \end{split}
\end{align}
As previously discussed, all the pertinent characteristic parameters, namely $\omega_m^{\rm nl}$ (bare frequency corrected by the nonlinear effect), $\alpha_m$ (qubit anharmonicity), and $g_{mn}$ (the coupling strength), can be efficiently determined through the IEPR method.
\par Furthermore, it is essential to consider the Hamiltonian in the laboratory frame and solve for the parameters of normal mode as well. Combing \eref{eq:ham_cir_tilde} and \eref{eq:ham_quantum}, the many-body bare Hamiltonian involving the nonlinear effect is expressed as
\begin{equation}
\label{eq:ham_bare_nl}
    \begin{split}
        \hat H^{\rm nl}_{\rm bare} &\simeq \sum_m \left[ \frac{ \hat{\tilde{\Phi}}_m^2}{2L_m} + \omega_m^2 \frac{\hat{\tilde{Q}}_m^2}{2} - \frac{L_m^J  (2\pi/ \Phi_0)^2}{24} \hat{\tilde{\Phi}}_m^4 \right ] \\
        & +\sum_{m\neq n}g_{mn} \sqrt{\omega_m \omega_n} \hat{\tilde{Q}}_m \hat{\tilde{Q}}_n.
    \end{split}
\end{equation}
In the equation above, $L_m^J$ represents the effective linear inductance of the Josephson junction for the $m$th element; for those linear elements, we set the corresponding $L_m^J$ to zero.
In line with the theory applied in the linear regime, the same transformation is employed on the nonlinear Hamiltonian, i.e., \eref{eq:ham_bare_nl}. In particular, the flux operator under bare-mode representation is expanded into a linear superposition of operators under normal-mode representation: $\hat{\tilde{\Phi}}_n = \sum_k u_{kn} \hat{\tilde{\Phi}}_k^\prime$. Therefore, the linear part of the Hamiltonian $\hat H^{\rm nl}_{\rm bare}$ is diagonalized, while the nonlinear part has to rewritten in terms of the normal-mode operators as well. Ultimately, it reads
\begin{equation}
\label{eq:ham_normal_1stQ}
\begin{split}
    \hat H^{\rm nl}_{\rm normal} &=\sum_m \hat{\tilde{\Phi}}_m'^2/2 + \omega_m'^2 \hat{\tilde{Q}}_m'^2/2 \\
    &-\sum_m \frac{L_m^J (2\pi/ \Phi_0)^2}{24} \left(\sum_k u_{km}\hat{\tilde{\Phi}}_k^\prime \right)^{4}.
\end{split}
\end{equation}
Applying the second quantization (the same as we performed in the linear regime) to the above equation yields the final nonlinear Hamiltonian
\begin{align}
    \label{eq:many_disp}
    \begin{split}
        \hat H_{\rm normal}^{\rm nl} &= \sum_m \omega_m^{\prime \rm{nl}} \hat{a}_m^\dagger \hat a_m + \frac{\alpha'_m}{2} \hat a_m^{\dagger 2} \hat a_m^2 \\
        &+ \sum_{m\neq n}\frac{\chi_{mn}}{2} \hat a_m^\dagger \hat a_m \hat a_n^\dagger \hat a_n.
    \end{split}
\end{align}
In the equation above, $\omega_m'^{\rm nl} $ denotes the renormalized-mode frequency, $\alpha'_m$, $\chi_{mn}$ are the self-Kerr and cross-Kerr parameters respectively. All of these crucial parameters are highly related to the experimental data. Using  \eref{eq:ham_normal_1stQ} and \eref{eq:many_disp}, they are calculated as follows
\begin{align}
    \label{eq:chi}
    &\chi_{mn} = 2\sum_k u_{mk}^2 u_{nk}^2 \frac{\omega_m^{\prime}\omega_n^{\prime}}{\omega_k^2} \alpha_k,\\
    \label{eq:alpha_m}
    &\alpha_m^\prime = \sum_k u_{mk}^4 \frac{\omega_m^{\prime 2}}{\omega_k^2} \alpha_k = \frac{\chi_{mm}}{2},\\
    \label{eq:renormalized_frequency}
    &\omega_m'^{\rm nl} =\omega_m' + \frac{1}{2}\sum_{k} \chi_{mk}.
\end{align}
These nonlinear parameters can also be determined using other existing methods, e.g., the BBQ \cite{nigg2012black} and EPR \cite{minev2021energy} methods. We will compare and discuss the results obtained with different methods in Sec.~\ref{subsection:nonlinear}.
It is evident that beyond the linear regime, the IEPR method can also be extended to the nonlinear regime. Our approach is versatile, capable of solving both linear and nonlinear Hamiltonian in  bare- and normal-mode representation.

\section{Procedure for quantum chip simulation and verification} 
\label{sec:procedure}
\begin{figure}[htbp]
    \centering
    \includegraphics[width=\linewidth]{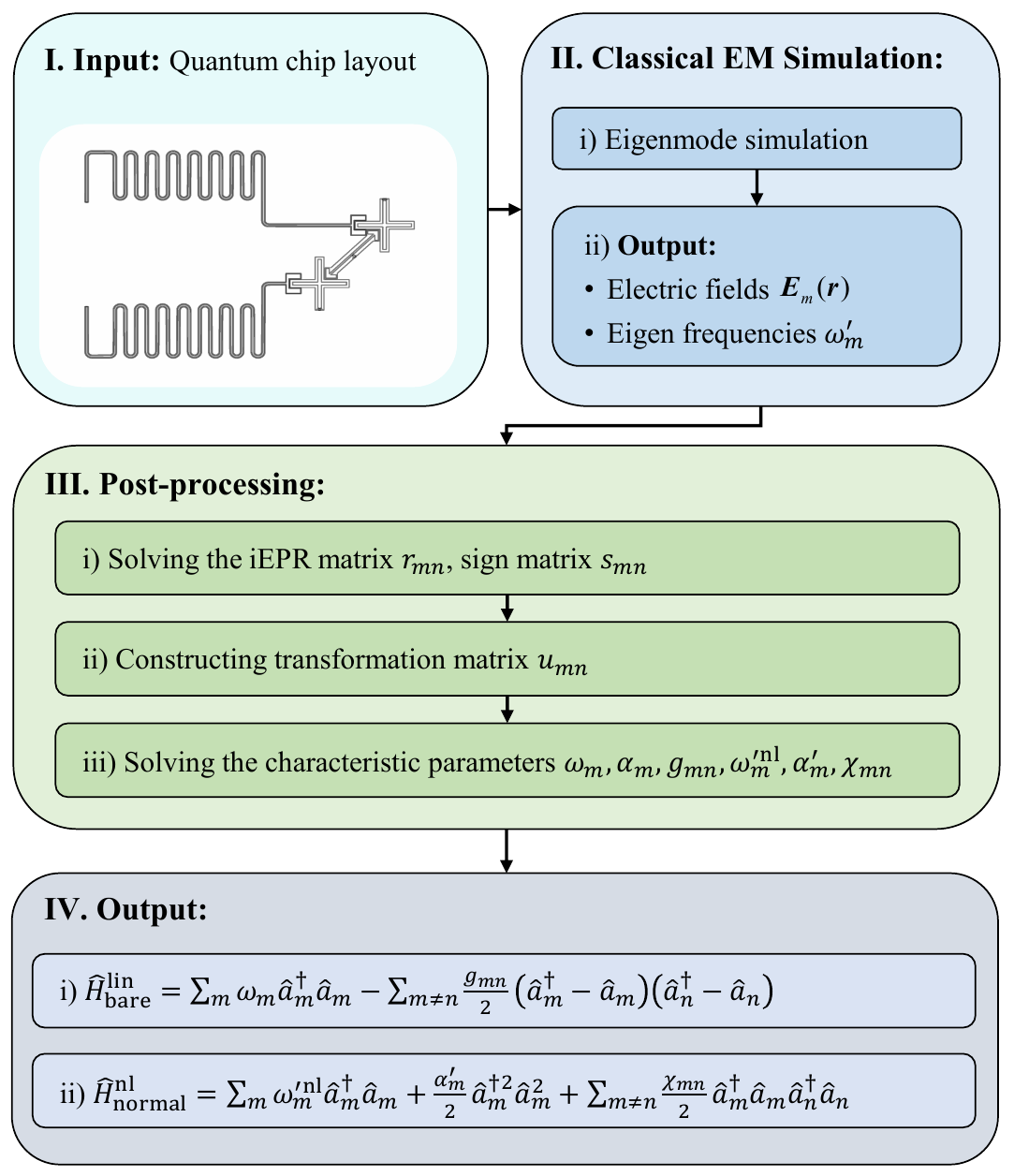}
    \caption{The procedure for quantum chip simulation and verification. Following this workflow, the desired Hamiltonian of superconducting quantum chip layouts is generated automatically and efficiently.  }
    \label{fig:workflow}
\end{figure}

Building upon the IEPR method, we have developed a comprehensive procedure designed to guide quantum chip simulation and verification. In particular, the procedure (see Fig.~\ref{fig:workflow}) commences from a given superconducting quantum chip layout and culminates in the determination of vital characteristic parameters and the Hamiltonian of interest. In particular, the procedure can be dissected into two principal segments: classical EM simulation and postprocessing. To embark on this journey, the classical EM simulation entails conducting eigenmode simulation, achieved by solving the three-dimensional Helmholtz equation within the framework of the quantum chip model. Ultimately, it outputs eigenmode frequencies $\omega_m'$ and the electric field $\bm{E}_m(\bm{r})\cos(\omega_m' t)$ corresponding to each normal mode $m$. Here, $\bm{E}_m(\bm{r})$ signifies the amplitude (peak value) of the electric field at position $\bm{r}$, encapsulating the insights garnered from the EM simulation.

\par Next, let us transition into the postprocessing part. It deals with the essential task of how to extract the characteristic parameters of interest from the classical EM simulation. Recalling the definition of the IEPR, i.e., \eref{eq:def_IEPR}, it necessitates the calculation of the average inductive energy for both the entire quantum chip and the specific elements. Notably, evaluating the inductive energy encompassed by the entire chip equates to gauging the average electric field energy \cite{minev2021energy}. This can be achieved through the integration of the electric field's spatial distribution over the entirety of the quantum chip's volume. Mathematically, this integration is encapsulated by $\mathcal E_m^{I} = \mathcal E_m^{E} = \int_{v_{\rm full}} [\epsilon(\bm{r}) |\bm{E}_m(\bm{r})|^2/4] dv$, where $\epsilon(\bm{r})$ signifies the permittivity at position $\bm{r}$. 
However, when it comes to determining the local inductive energy of a specific element, complexities arise.
To address this challenge, we adopt an alternative approach: the inductive energy exhibits a linear correlation with the power of flux passing through the element.
Consequently, the average inductive energy encapsulated within the specific elements is quantified as $\mathcal{E}_{mn}^I = \Phi_{mn}^2/4L_n = V_{mn}^2/(4L_n \omega_m'^2)$ (see Appendix \ref{sec:solve_IEPR} for more details). Here, $\Phi_{mn}$ corresponds to the peak value of the flux, and $V_{mn}$ represents the peak voltage across the $n$-th inductor of the element under mode $m$. Notably, the peak voltage is evaluated as $V_{mn} = \int_{\bm{l}_n}\bm{E}_m (\bm{r}) d\bm{l}$, where the integration path $\bm{l}_n$ is situated between two potential nodes along the element (see Appendix \ref{sec:solve_IEPR} for more details). 
By employing \eref{eq:def_IEPR}, the IEPR materializes as $r_{mn} = [V_{mn}^2/(4L_n \omega_m'^2)]/\mathcal E_m^{I}$. As for those cases that do not own clear potential nodes, e.g., three-dimensnal (3D) resonator or parasitic box modes, IEPR method still holds in normal-mode representation. The relevant discussion is postponed in Sec.~\ref{subsection:others}.
Through the orthonormality property established for the IEPR matrix, $\sum_m r_{mn}=\sum_m[V_{mn}^2/(4L_n \omega_m'^2)]/\mathcal E_m^{I}= 1$, we unravel the phenomenological parameter $L_n = \sum_m(V_{mn}^2/ \omega_m'^2)/4\mathcal E_m^{I}$ and, subsequently, the IEPR is
\begin{align}
\label{eq:IEPR_sim}
r_{mn} = \frac{V_{mn}^2/ (\omega_m'^2 \mathcal{E}_m^I)}{\sum_{k}V_{kn}^2/ (\omega_k'^2 \mathcal{E}_k^I)}.
\end{align}
In the denominator of the equation above, $k$ represents the index of different modes, and the summation includes the IEPR values across these modes for the $n$th element. Since $L_n$ is a phenomenological parameter stemming from the energy distribution, the IEPR method exhibits versatility and can be extended to all inductive elements, in contrast to the EPR method \cite{minev2021energy} which is confined to a scenario where the element's inductance has to be known.
As a further step, $s_{mn}$ can be determined by introducing a reference-oriented line segment and determining the electric field distribution. We can ascertain $s_{mn}$ by checking the sign of $V_{mn}$. More details can be found in Appendix \ref{sec:solve_IEPR}. Combining with $r_{mn}$ and $s_{mn}$, we are able to reconstruct the transformation matrix $U$.
This lays the groundwork to derive the bare frequencies and coupling strengths within the linear regime, facilitating the determination of $\hat H_{\rm bare}^{\rm lin}$. Furthermore, as we venture into the nonlinear regime, the bare-mode frequency adapts to $\omega_m^{\rm nl}$, and the anharmonicity $\alpha_m$ is likewise evaluated. These parameters collectively pave the way for the emergence of the bare-mode Hamiltonian in the nonlinear regime, denoted as $\hat H^{\rm nl}_{\rm bare}$. For the normal-mode Hamiltonian involving nonlinear effects, denoted as $\hat H^{\rm nl}_{\rm normal}$, the relevant parameters can be aptly represented as a linear superposition of the parameters in the bare mode, a relationship explicitly captured by Eqs.~\eqref{eq:chi}-\eqref{eq:renormalized_frequency}, which are thoughtfully constructed through the IEPR methodology. 

\section{Applications and Discussions}
\label{sec:applications}
To substantiate the effectiveness and advantages of both the IEPR method and the proposed procedure, we apply them to various different applications. Our initial focus centers on the examination of linear coupling characteristics of the layout depicted in Fig.~\ref{fig:layout} in Sec.~\ref{subsection:linear}. Here, we successfully ascertain essential characteristic parameters, with a particular emphasis on obtaining coupling strengths across different coupling regimes.
Subsequently, we delve into the exploration of nonlinear characteristics in Sec.~\ref{subsection:nonlinear}. Our particular attention is directed toward the evaluation of renormalized-mode frequencies, self-Kerr and cross-Kerr parameters.
Moreover, we demonstrate the extensibility of the IEPR method to handle high-order resonator modes and the coupling between qubits and the three-dimensional resonator in Sec.~\ref{subsection:others}. Comprehensive analyses and discussions regarding these aspects are presented.
In conclusion, Sec.~\ref{subsection:features} encompasses a summation of various different methods employed for the study and validation of the IEPR approach.

\subsection{The linear coupling characteristics}
\label{subsection:linear}
To construct the bare Hamiltonian  of the layout depicted in Fig.~\ref{fig:layout} in the linear regime, several key parameters need to be determined: i) the bare frequencies of qubit, coupler and resonator; ii) the dispersive-type coupling strength between qubit and resonator; iii) the dispersive-type coupling strength between qubit and coupler; iv) the resonant-type direct coupling strength between two qubits. 
Here, the ``direct coupling" represents the mutual capacitive coupling between the qubits.
\par
\begin{figure}
    \centering
\includegraphics[width=\linewidth]{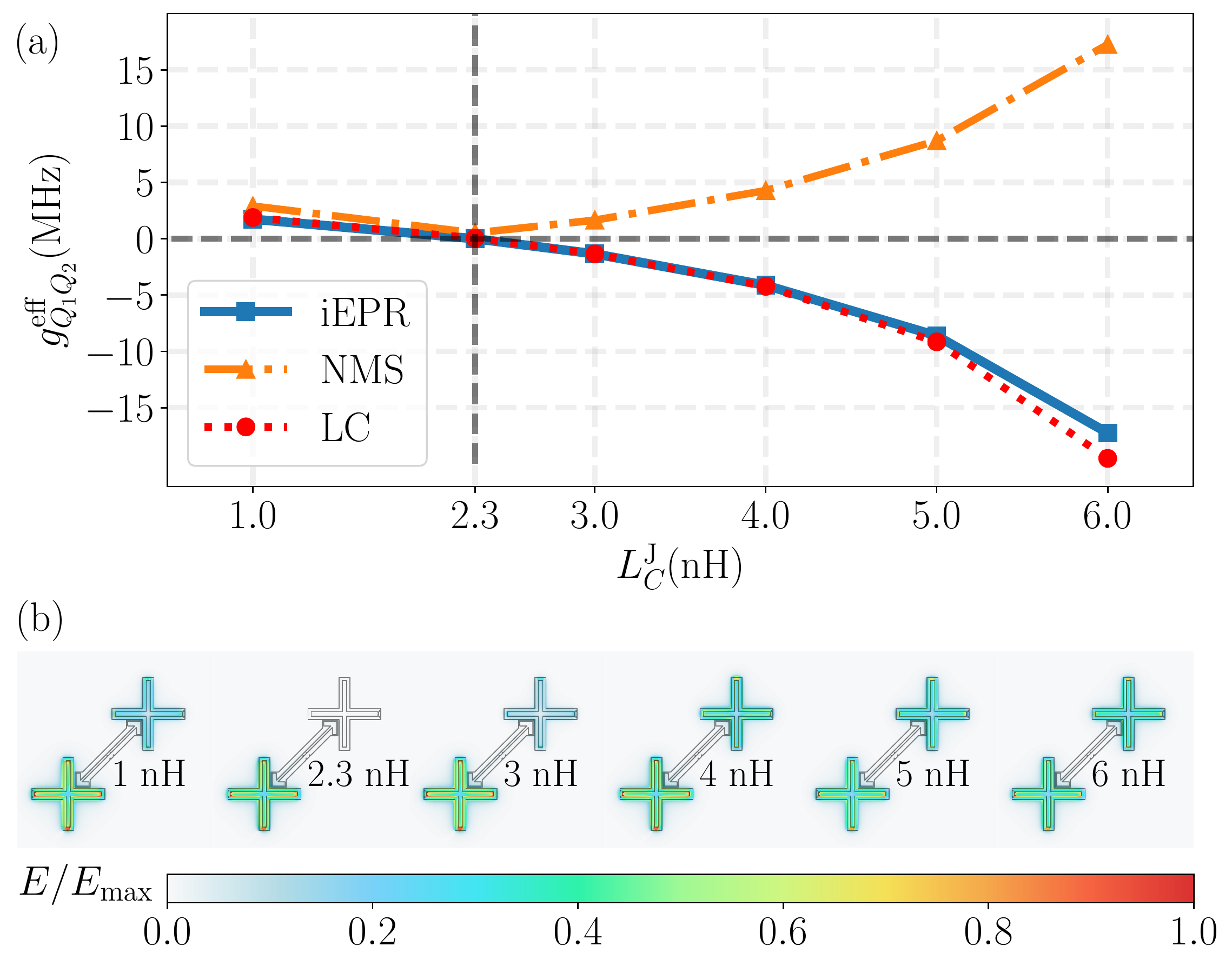}
    \caption{(a) Qubit-qubit effective coupling characteristics. With varying the coupler inductance $L_C^J$, the effective coupling strength $g_{Q_1Q_2}^{\rm eff}$ is tuned from positive to negative, and across zero around $2.3 \, \rm{nH}$. The results obtained using the IEPR method (blue solid) are verified simultaneously using the NMS method (orange dash-dotted) and the LC method (red dotted).  (b) The electric field distribution $E$ (normalized by the maximum $E_{\rm max}$)  of the qubit mode corresponding to different $L_C^J$ in (a).}
\label{fig:Applications_figure}
\end{figure}

 The procedure outlined in Fig.~\ref{fig:workflow} is systematically applied to address these pertinent parameters in the layout. Our journey begins with the initiation of classical electromagnetic (EM) simulations. The quantum chip layout is modeled in 3D using ANSYS HFSS, with comprehensive details provided in Appendix \ref{sec:simMethod}. The layout comprises various intricate components, including qubits, couplers, resonators, and a grounding plane. The superconducting metal layer boasts ideal conductivity and spans dimensions of $10.5 \times 10.5$ $\rm{mm}^2$. This layout resides atop a sapphire substrate featuring a thickness of 1 $\rm{mm}$ and a relative permittivity of 10.5. Enclosed within a vacuum box reaching a height of 29 mm, the complete 3D model embodies intricate spatial features and material characteristics. The culmination of this simulation yields the crucial characteristic parameters, which are then presented in Tab.~\ref{tab:full}. Observing the data, we note that the diagonal entries correspond to the bare frequencies of the qubit, coupler, and resonator, while the off-diagonal entries represent the coupling interactions between different elements.
With these determined parameters in hand, we are empowered to construct the bare Hamiltonian for the specific layout under investigation, thus enabling us to explore its dynamic evolution.
\begin{table}[b]
\centering
\begin{ruledtabular}
    \begin{tabular}{cccccc}
         &$Q_1$ &$Q_2$ &$C$ &$R_1$ &$R_2$\\
        \hline 
        $Q_1$ &$6699.371$ &$8.76$ &$158.05$ &$40.08$ &$1.33$ \\
        $Q_2$ &$8.76$ &$6701.12$ &$159.92$ &$0.37$ &$40.56$ \\
        $C$  &$158.05$ &$159.92$ &$8118.31$ &$1.68$ &$14.00$\\
        $R_1$ &$40.08$ &$0.37$ &$1.68$ &$5019.11$ &$0.36$ \\
        $R_2$ &$1.33$ &$40.56$ &$14.00$ &$0.36$ &$5092.84$
    \end{tabular}
\end{ruledtabular}
\caption{The characteristic parameters (unit: MHz) of the layout of interest solved using the procedure.  The diagonal terms represent the bare frequency of each element, while the off-diagonal terms describe the coupling between different elements.  The inductance of qubits and coupler are taken as 6.691 nH, 5.408 nH respectively.}
\label{tab:full}
\end{table}
In addition to the critical parameters, another significant endeavor is the characterization of qubit-qubit effective coupling, which is the combination of the direct coupling referred and the indirect coupling induced by the tunable coupler \cite{yan2018tunable}.
Within the coupler architecture, a noteworthy capability lies in the tunability of the effective coupling between neighboring qubits. In the context of electromagnetic simulation, achieving this tunability is realized through the manipulation of the coupler frequency, effectively adjusting the effective inductance of the coupler itself.  Although the layout encompasses a total of five elements, the intricate consideration of all potential nodes, which is often a requirement in the lumped-circuit (LC) method, is bypassed. Instead, we strategically narrow our focus to the two modes directly relevant to the qubits. This simplification permits a convenient solution for the effective coupling, and the rationale behind this reduction from a many-body to an effective two-body system is validated in Appendix \ref{sec:reduce}.
Once again, employing the step-by-step process illustrated in Fig.~\ref{fig:workflow} and varying the coupler inductance $L_C^{\rm J}$, we successfully solve and presente the qubit-qubit effective coupling characteristics in Fig.~\ref{fig:Applications_figure}(a). The plot demonstrates that the effective coupling strength, as predicted by the IEPR method, transitions from positive to negative as the coupler inductance increases, and reaching a point of coupling ``switch-off" around $2.3~{\rm nH}$. For validation, we compare the IEPR results (blue solid) with two existing methods: NMS (orange dash-dotted) and LC (red dotted). Regrettably, both the BBQ and EPR methods are unsuitable for addressing this crucial issue. The congruence between the IEPR approach and the other two methods underscores the effectiveness of the IEPR method and the accompanying procedure for quantum chip simulation and verification. Next, let us now delve into some pertinent observations regarding these three distinct methodologies. The NMS method encounters difficulty in determining the sign of the coupling strength, rendering the validation of the ``switch-off" property of the coupler architecture a difficult task. On the other hand, as the number of integrated qubits on the quantum chip increases, the LC method faces substantial challenges. In contrast, the IEPR method remains resilient in the face of these obstacles and demonstrates pronounced advantages.

Last but not least, we have made an intriguing observation regarding the interplay between the electric field distribution of the qubit mode and the qubit-qubit effective coupling characteristics. As shown in Fig.~\ref{fig:Applications_figure}(b), when the electric field is localized on a single qubit, as seen at $2.3 \, {\rm nH}$, it signifies that the qubit remains isolated and does not interact with its counterparts, thus aligning with the ``switch-off" point of zero coupling. As a comparison, when the electric field extends across both qubits, a strong effective coupling emerges, indicating robust interaction between the qubits. 
This correspondence underscores the profound relationship between inductive energy participation and the resulting coupling characteristics. Indeed, this intuitively insightful discovery serves as the primary impetus for the development of this work. Eventually, our work not only furnishes a comprehensive theoretical elucidation of this  correspondence but also yields  quantifiable results, enhancing our understanding of these intricate interactions.

\subsection{The nonlinear characteristics: self-Kerr and cross-Kerr parameters}
\label{subsection:nonlinear}

In addition to the linear coupling characteristics, the nonlinear effects arising from Josephson junctions play a pivotal role in the behavior of superconducting quantum chips. Concentrating on the nonlinear effects, the IEPR method becomes instrumental in predicting the renormalized mode frequency, as well as the self-Kerr and cross-Kerr parameters, which hold significant relevance to experimental observations. To illustrate the power of the IEPR method in capturing the nonlinear characteristics, we focus on a specific subsystem comprising of qubit $Q_1$ and coupler $C$ within the layout shown in \fref{fig:layout}.

\begin{table*} 
\centering
\begin{ruledtabular}
    \begin{tabular}{cccccc}
        &\textbf{IEPR[This work]} & LC\cite{yurke1984quantum,van1997conductance,burkard2004multilevel,malekakhlagh2017cutoff,gely2020qucat,minev2021circuit} &BBQ\cite{nigg2012black,bourassa2012josephson,solgun2014blackbox} &EPR\cite{minev2021energy}   \\
        \hline 
        $\omega_q^{\prime \rm nl}$  & $6439.47$ &$6545.06$ &$6445.32$ &$6442.31$  \\
        $\omega_c^{\prime \rm nl}$  & $7817.38$ &$7962.57$ &$7827.60$ &$7821.22$   \\
        $\alpha_q^\prime$ & $-222.42$ & $-231.11$ &$-227.61$ &$-220.01$   \\
        $\alpha_c^\prime$ & $-261.95$ & $-270.22$ &$-266.01$ &$-258.53$ & \\
        $\chi_{qc}$ & $-14.01$ & $-14.84$ &$-13.77$ &$-13.17$ \\
        \hline
        Time consumption(s) & $740.23$ & $192.35$ &$1244.86$ &$740.23$ \\
         \hline\hline
        Bare Hamiltonian & \Checkmark  & \Checkmark  & \XSolid  &\XSolid  \\
        Normal Hamiltonian &\Checkmark &\Checkmark &\Checkmark &\Checkmark \\
        3D resonator &\Checkmark &\XSolid &\Checkmark &\Checkmark
    \end{tabular}
\end{ruledtabular}
\caption{ 
Comparison of various quantum chip simulation and verification methods employed in the subsystem $Q_1-C$ of the layout shown in \fref{fig:layout}. IEPR, LC, BBQ and EPR methods are applied to solve the nonlinear parameters (unit: MHz) in normal-mode representation. Time consumption is  measured with the same computer to identify the efficiency of different methods. Comparing with existing methods, IEPR method exhibits broader applicability to resolve various type of tasks.}
\label{tab:comp}
\end{table*}

First of all, we specify the electric parameters used in our simulation: the inductance for qubit $Q_1$ and coupler are 6.691 nH, 5.4 nH, respectively. The inductance for qubit $Q_2$ is set as 0.1 nH, intentionally detuned from $Q_1$ and $C$ to eliminate the parasitic coupling effect.
Through utilizing the procedure outlined in \fref{fig:workflow}, we comprehensively determine the renormalized frequencies of the qubit and coupler, denoted as $\omega_q'^{\rm nl}$ and $\omega_c'^{\rm nl}$, respectively. Moreover, we extract the self-Kerr parameters $\alpha_q'$ and $\alpha_c'$ for the qubit and coupler, along with the cross-Kerr parameter  $\chi_{qc}$ characterizing their dispersive coupling. These results are summarized in the second column of Tab.~\ref{tab:comp}, providing a clear representation of the quantitative insights enabled by the IEPR method. 
The correctness of these nonlinear parameters has been verified using a complementary numerical technique (see technique details in Appendix \ref{sec:simMethod}). 
Additionally, to ensure robustness and comprehensiveness, our analysis encompasses a comparison with various well-established existing methods, all of which are detailed in Tab.~\ref{tab:comp} (refer to Appendix \ref{sec:simMethod} for the specific techniques employed). Impressively, our findings underscore the remarkable agreement between the solutions obtained via the IEPR method and those solved from conventional EPR \cite{minev2021energy}, BBQ \cite{nigg2012black,bourassa2012josephson,solgun2014blackbox}, and LC \cite{yurke1984quantum,van1997conductance,burkard2004multilevel,malekakhlagh2017cutoff,gely2020qucat,minev2021circuit} methods. This harmony across diverse methodologies attests to the versatility and reliability of the IEPR method when addressing intricate nonlinear characteristics.

\subsection{Other crucial characteristics}
\label{subsection:others}

Next, we  turn attention to some additional considerations within the realm of quantum chip characteristics. It is pertinent to acknowledge the existence of higher-order modes, including those associated with resonators, structural parasitic modes, as well as box modes. 
Employing the IEPR method, we can also estimate quantitatively the impact or contribution induced by these modes. 
An illustrative example involves assessing the coupling strength between the qubit and the higher-order resonator mode depicted in \fref{fig:layout}. While the coupling between the fundamental resonator mode and the qubit has been documented in \tref{tab:full}, the higher-order resonator mode could potentially exert influence on the qubit as well. If the dissipation of a particular mode is substantial, it can significantly compromise the quality factor of the qubit \cite{houck2008controlling}. Consequently, we conduct EM simulations on the higher-order resonator mode, e.g., the third-order mode of readout resonator. Following the procedure introduced in \sref{sec:procedure}, we compute the IEPR of this higher-order mode and perform postprocessing in a similar way. At the end, we obtain that the third-order modes of $R_1$ with a frequency of $15$ GHz and $R_2$ with a frequency of $15.1$ GHz exhibit coupling strengths of 68.45 and 68.06 MHz with their respective qubits.

\begin{figure}[t]
    \centering
    \includegraphics[width=\linewidth]{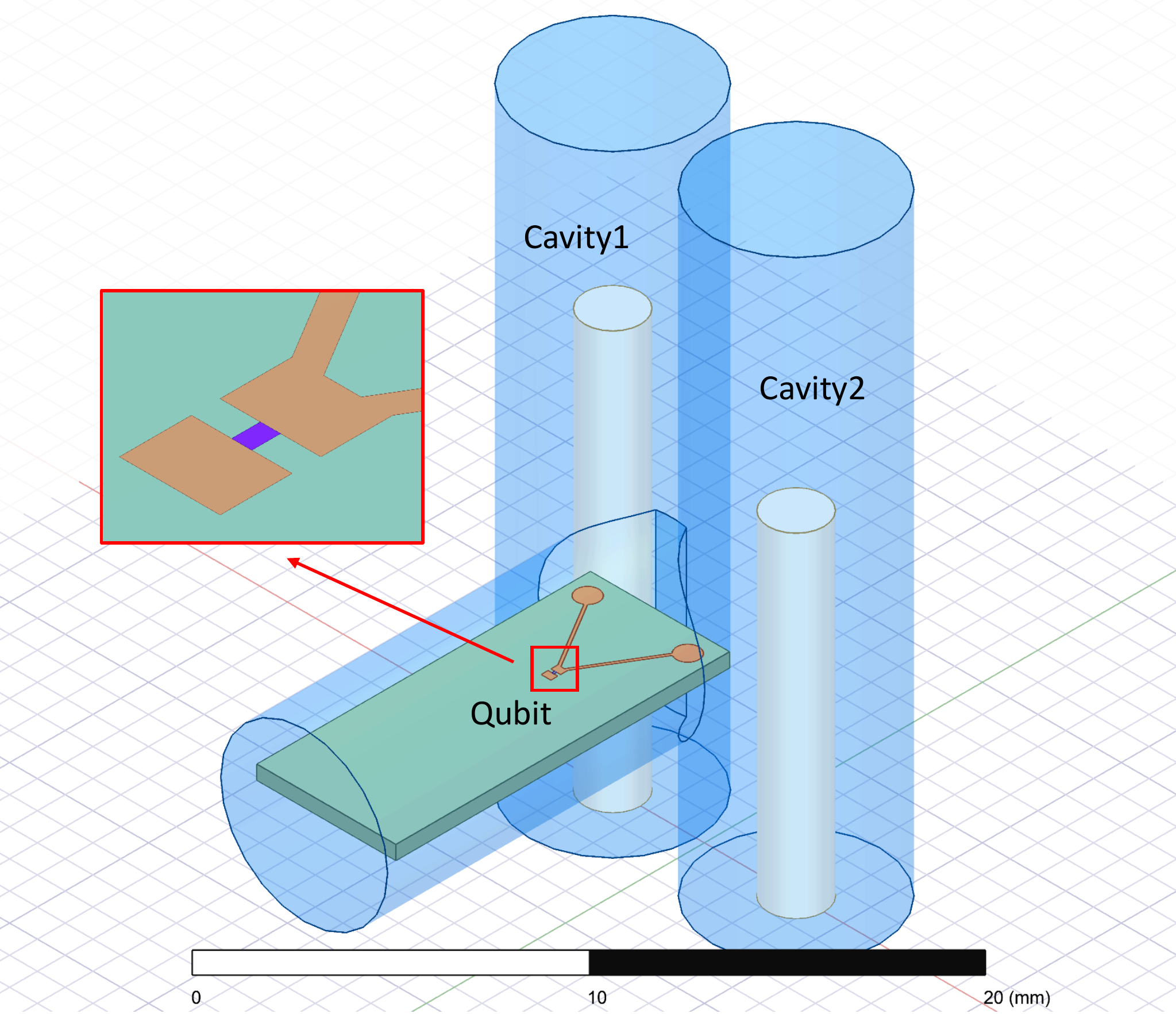}
    \caption{A three-dimensional schematic of the system consisting of two 3D cavities and a single transmon qubit. 
    Both of the  cavities (labeled as ``Cavity1" and ``Cavity2") are coupled to the qubit. 
    The cavities (blue) are superconductors with different central axes (white) height. Inset: the view of qubit. The Josephson junction (purple) is sandwiched between two superconducting pads (orange), and the structure is on the surface of the sapphire substrate (green). IEPR method is applied to solve the characteristic parameters of interest.} 
    \label{fig:transmon_3Dcavity}
\end{figure}

Referring to the techniques explained in \sref{sec:procedure}, the pivotal aspect in solving IEPR lies in computing the voltage across the two potential nodes of an element, facilitating IEPR determination through orthonormality. However, when potential nodes are less distinct, as observed in scenarios like higher-order modes with multiple nodes or complex 3D resonator, the task becomes more challenging. 
Consequently, the linear coupling strengths associated with these modes cannot be computed using Eqs.~\eqref{eq:baref} and \eqref{eq:coup}. Nonetheless, nonlinear parameters like self-Kerr and cross-Kerr can still be determined. Notably, Kerr nonlinear parameters solely rely on columns corresponding to qubits with discernible potential nodes, allowing for the acquisition of IEPR under different modes. These columns can be normalized to construct the corresponding columns in the $U$ matrix, enabling the complete resolution of nonlinear parameters of any modes. However, establishing bare-mode frequency and coupling strength remains a challenge. Alternatively, if collecting inductive energy proves impractical, a technique leveraging the orthonormality of the unitary matrix has been devised to address the issue of a single mode without clear potential nodes. Leveraging this insight, if the first n-1 columns of the matrix are definitively determined, the last column can be unambiguously derived through orthogonality principles. That is to say, this approach allows one element within the system not to require IEPR calculation. Consequently, both linear and nonlinear characteristics parameters of an $n$-body system containing $n-1$ elements with clear potential nodes and a single element devoid of potential nodes can be solved. 
Lastly, we clarify the limitation of this method. If one is interested in solving the linear coupling characteristics of the scenarios where more than one mode is without clear potential nodes, the method would be no longer applicable.

\begin{table*}[t]
\centering
\begin{ruledtabular}
    \begin{tabular}{cccccccccc}
     $\hat{H}^{\rm nl}_{\rm bare}$&$\omega_q$ &$\omega_{1}$ &$\omega_{2}$  &$\alpha_q$ &$\alpha_{1}$ &$\alpha_{2}$ &$g_{q1}$ &$g_{q2}$ &$g_{12}$ \\
    \hline
    IEPR &5068.88 &4662.27 &5742.23 &-255.46  &0 &0 &16.62~(16.67) &25.01~(25.41) &N/A\\
    \hline \hline
     $\hat{H}^{\rm nl}_{\rm normal}$&$\omega_{q}^{\prime \rm{nl}}$ &$\omega{1}^{\prime \rm{nl}}$ &$\omega_{2}^{\prime \rm{nl}}$ &$\alpha_{q}^{\prime}$ &$\alpha_{1}^{\prime}$ &$\alpha_2^\prime$ &$\chi_{q1}$ &$\chi_{q2}$ &$\chi_{12}$ \\
    \hline
    IEPR &4814.22 &4661.87 &5741.83 &-253.87 &-6.01E-4 &-6.18E-4 &-0.78 &-0.79 &-1.21E-3\\
    EPR &4827.02 &4661.89 &5741.85 &-241.11 &-5.69E-4 &-5.85E-4 &-0.74 &-0.75 &-1.15E-3\\
    BBQ &4836.72 &4665.01 &5762.36 &-244.81 &-5.45E-4 &-5.65E-4 &-0.72 &-0.74 &-1.11E-3
    \end{tabular}
\end{ruledtabular}
\caption{The solved characteristic parameters (unit: MHz, in both bare- and normal-mode representations) of the system displayed in Fig.~\ref{fig:transmon_3Dcavity}. The linear couplings between the qubit and cavities are solved and verified with the NMS method, using which the results are obtained in parentheses. The parameters in normal modes are solved using the IEPR, EPR and BBQ method, respectively.}
\label{tab:transmon_3Dcavity}
\end{table*}

To verify the techniques introduced above, we apply them to a scenario involving double 3D cavities coupled with a single qubit (depicted in Figure \ref{fig:transmon_3Dcavity}) where both of the cavities do not have clear potential nodes. This architecture is studied for realizing two cat states entanglement \cite{wang2016schrodinger} and dual-rail encoding \cite{teoh2023dual}, aims to explore the scalability of 3D cavity architecture. Specifically, the configuration involves a transmon qubit interconnected via a Josephson junction to two pads on the top surface of a sapphire substrate. The chip measures $12 \times5 \times0.5~{\rm mm}^3$, with a sapphire substrate having a relative permittivity of 10.5, and the qubit's inductance is 13 nH. The pads connected to the junction ends exhibit perfect conductor boundary conditions. Enclosed within a vacuum cylindrical cavity package with a radius of 3 mm and a height of 13.2 mm, the qubit chip interacts with two 3D cavities labeled as ``Cavity1" and ``Cavity2", each possessing a radius of 3 mm and a height of 22 mm. The inner axis of the 3D cavities consists of cylindrical superconductors with the same radius of 1 mm and different heights (15 mm and 12 mm) to avoid frequency collision.
Following the techniques explained, the IEPR for the qubit is solved benefiting from the clear potential nodes. Subsequently, leveraging orthonormality principles, the crucial transformation matrices  $U$  for the subsystems ``Qubit-Cavity1" and ``Qubit-Cavity2" are constructed. Through the procedure detailed in Sec.~\ref{sec:procedure}, we are able to extract most of the characteristic parameters (e.g., the bare-mode frequency $\omega_{q,1,2}$, the anharmonicity $\alpha_{q,1,2}$, and the couplings $g_{q1,q2}$) in bare-mode representation, except the linear coupling $g_{12}$ between the two cavities because they do not own clear potential nodes. Additionally, owing to the solvability of IEPR for various modes on the qubit,  the corresponding column of the qubit in the $U$ matrix can also be solved. This facilitates the resolution of nonlinear parameters such as renormalized frequency (e.g., $\omega'^{\rm nl}_{q,1,2}$), self-Kerr parameters (e.g., $\alpha'_{q,1,2}$), as well as cross-Kerr parameters (e.g., $\chi_{q1}$, $\chi_{q2}$, $\chi_{12}$) with the aid of Eqs.~(\ref{eq:chi})-(\ref{eq:renormalized_frequency}). The resultant characteristic parameters in both bare- and normal-mode representations are calculated and summarized in Table \ref{tab:transmon_3Dcavity}. While the EPR and BBQ methods are not applicable for bare-mode representation, we employ the NMS method to cross-verify the linear coupling strengths between the qubit and 3D cavities. It is clearly shown that the NMS results (correspond to the values in the parentheses of the second row)  are in good agreement with IEPR results, more technical details concerning the NMS method are presented in Appendix \ref{sec:simMethod}. Besides, we also notice that the method is invalid for solving the linear coupling between two cavities. As for the critical characteristic parameters in normal-mode representation, we demonstrate them using both EPR and BBQ methods simultaneously, showcases the consistency of IEPR results.
\par While our attention has thus far centered around ideal closed systems, it is crucial to consider the real-world scenario where superconducting quantum chips are coupled to external environments, leading to dissipation. This phenomenon is well illustrated by a readout module \cite{blais2021circuit} where qubits are coupled to resonators that are in turn directly coupled to transmission lines. 
Given the signal's input and output role, the transmission line acts as an external dissipation source, causing photon loss in both qubit and resonator. The loss rate of the resonator coupled with the transmission line is correlated with the coupling strength, defined as $\kappa_r = Z_t \omega_r^2 C_{rt}^2/C_r = 4g_{rt}^2/\omega_r$ \cite{blais2021circuit}, while the loss rate of the qubit isolated by the resonator is determined by $\gamma_q = (g_{qr}/\Delta_{qr})^2 \kappa_r$ \cite{blais2021circuit, krantz2019quantum,houck2008controlling}. Obviously, it depends on various parameters including $Z_t$, the characteristic impedance of the transmission line, $\omega_r$, the bare frequency of the resonator, $C_{rt}, C_{r}$, the mutual capacitance between the resonator and the transmission line and the self-capacitance of the resonator respectively, $g_{rt}, g_{qr}$, the coupling strengths of the resonator and the transmission line, and the resonator and the qubit respectively, $\Delta = \omega_q - \omega_r$, the detuning of qubit frequency with respect to the resonator. Notably, the losses are linked to coupling strengths. Since IEPR method can be applied to solve coupling strengths among different elements, it allows us to further calculate loss rates.

\subsection{Features of the IEPR method}
\label{subsection:features}
In a comparative analysis with numerous existing methods \cite{yurke1984quantum,van1997conductance,burkard2004multilevel,malekakhlagh2017cutoff,gely2020qucat,minev2021circuit,nigg2012black,bourassa2012josephson,solgun2014blackbox,minev2021energy,dubyna2020inter}, the IEPR method emerges with advantages for characterizing superconducting quantum chips.
It exhibits adaptability to different coupling regimes and the ability to ascertain relative positive and negative coupling strengths. Moreover, IEPR method delivers comprehensive characteristic parameters, encompassing both linear and nonlinear parameters in bare- and normal-mode representations. This versatility distinguishes it as an adaptable tool in the realm of superconducting quantum chip analysis.
A notable strength of the IEPR method lies in its ability to resolve couplings not only between chip elements but also with additional modes, including complex three-dimensional resonators and certain parasitic modes that pose challenges for conventional lumped-circuit methods. This capability enhances the method's scope and applicability, enabling a more comprehensive assessment of quantum chip behavior in real-world scenarios.
Conversely, other widely employed methods exhibit their own distinct attributes and limitations. For instance, the lumped-circuit (LC) method \cite{yurke1984quantum,van1997conductance,burkard2004multilevel,malekakhlagh2017cutoff,gely2020qucat,minev2021circuit}, which relies on static EM simulation, renowned for its efficiency and suitability for solving bare and normal parameters. However, this method treats continuous conductors as equipotential bodies, resulting in the loss of significant high-frequency field information. Consequently, its simulation outcomes are approximations, effective primarily for lumped elements significantly smaller than the wavelength, but notably less accurate for distributed parameter elements. Moreover, the LC method encounters challenges when dealing with elements lacking specific potential nodes, such as 3D resonators and parasitic modes, remain unsolved.
The black-box quantization (BBQ) method \cite{nigg2012black,bourassa2012josephson,solgun2014blackbox} relies on full-wave simulations, making it suitable for high-frequency electromagnetic field scenarios. It can effectively derive parameters like self-Kerr and cross-Kerr in normal-mode representation. Nevertheless, this technique hinges on driven modal simulations, which, albeit slightly slower, are still a valuable tool. Unfortunately, BBQ falls short in handling bare-mode parameters.
Similarly, the energy participation ratio (EPR) method \cite{minev2021energy} shares many parameters and constraints with BBQ. It employs eigenmode simulations, offering improved computational speed compared to BBQ.
Conversely, the normal-mode simulation (NMS) method \cite{dubyna2020inter} is only suitable for resonance cases. 
To obtain the coupling characteristics, multiple full-wave simulations are required for sampling which is similar to the real experimental measurement. The whole process is inefficient, requiring numerous EM simulations but providing limited characteristics.
Remarkably, while most existing methods predominantly focus on parameters in normal modes, the IEPR method innovatively bridges the gap by simultaneously exploring characteristics in both bare-mode and normal-mode representations. This distinctive feature opens up alternative research avenues and provides a more comprehensive understanding of individual elements and their direct couplings, crucial for optimizing chip performance and designing efficient quantum chips.

\section{Summary and Outlook} 
\label{sec:summary}
In summary, our work introduces the IEPR method, along with a concise simulation and verification procedure, to effectively characterize superconducting quantum chips.
IEPR offers an efficient method for generating key characteristic parameters and Hamiltonians, as evidenced by our successful application in coupler architecture and 3D resonator scenarios.
Looking ahead, the IEPR method opens doors to exploring various challenges in quantum chip design. This includes investigating coupling phenomena between parasitic modes and element modes \cite{hornibrook2012superconducting}, addressing crosstalk issues \cite{mundada2019suppression, ku2020suppression, sung2021realization}, and delving into surface-loss problems \cite{wenner2011surface} may be studied using the IEPR method. 
By providing a well-defined procedure, our method represents a significant stride toward realizing quantum electronic design automation for superconducting quantum chips. This not only advances our understanding of quantum chip behavior but also paves the way for more efficient  quantum hardware development.

\begin{acknowledgments}
We would like to thank Yuan-hao Fu and Fei-Yu Li for their valuable discussion.
This work was done when K.Y. and X.J. were research interns at Baidu Research.
\end{acknowledgments}

\appendix

\section{The relation between IEPR and representation transformation}
\label{sec:relation_IEPR}
In this section, we delve into the intricate relationship between the IEPR and representation transformation.  Expressed in the vector space $\bm{\Lambda}$ of bare modes, the linear bare Hamiltonian (as shown in \eref{eq:ham_cir_tilde}) of a capacitively coupled system can be rewritten in the matrix form as
\begin{align}
    \label{eq:bareHam_mat}
    \hat{H}_{\rm{bare}} = \frac 1 2 \begin{pmatrix}
    1 &0 &\cdots &0\\
    0 &1 &\cdots &0 \\
    \vdots &\vdots &\ddots &\vdots\\
    0 &0 &\cdots &1\\
     & & & &\omega_1^2 &\tilde{g}_{12} &\cdots &\tilde{g}_{1n}\\
     & & & &\tilde{g}_{12} &\omega_2^2  &\cdots &\tilde{g}_{2n}\\
     & & & &\vdots &\vdots &\ddots &\vdots \\
     & & & &\tilde{g}_{1n} &\tilde{g}_{2n}  &\cdots &\omega_{n}^2 \\
    \end{pmatrix} = \frac 1 2 \begin{pmatrix}
        \mathcal{I} &0\\
        0 &\mathcal{H} 
    \end{pmatrix},
\end{align}
where we used $\tilde{g}_{mn} = 2g_{mn}\sqrt{\omega_m \omega_n}$ for simplicity. \eref{eq:bareHam_mat} is a real symmetric matrix that is diagonalizable. Expressed in the vector space $\bm{\Lambda}'$ of normal modes, the resulting normal Hamiltonian can be represented in matrix form as 
\begin{align}
    \label{eq:normHam_mat}
    \hat{H}_{\rm{normal}} = \frac 1 2 \begin{pmatrix}
    1 &0 &\cdots &0\\
    0 &1 &\cdots &0 \\
    \vdots &\vdots &\ddots &\vdots\\
    0 &0 &\cdots &1\\
     & & & &\omega_1'^2 &0 &\cdots &0\\
     & & & &0 &\omega_2'^2  &\cdots &0\\
     & & & &\vdots &\vdots &\ddots &\vdots \\
     & & & &0 &0  &\cdots &\omega_{n}'^2 
    \end{pmatrix}= \frac 1 2 \begin{pmatrix}
        \mathcal{I} &0\\
        0 &\mathcal{H}'
    \end{pmatrix}.
\end{align}
The equation above is derived through a unitary transformation,  specifically $\hat{H}_{\rm{normal}} = U\hat{H}_{\rm{bare}}U^\dagger$.  Since the bare Hamiltonian exhibits a block-diagonal structure, it necessitates that the transformation matrix $U$ also assumes a block diagonal format as
\begin{align}
    \label{eq:U}
    U = \begin{pmatrix}
        \mathcal{V} &0\\
        0 &\mathcal{U}
    \end{pmatrix}.
\end{align}
Considering the commutation relation must be conserved after the transformation, expressed as
\begin{equation}
\begin{split}
     \left[\hat{\tilde{\Phi}}_m^{\prime}, \hat{\tilde{Q}}_n^{\prime}\right]&=\sum_{kl}v_{m k} u_{n l} \left[  \hat{\tilde{\Phi}}_k,   \hat{\tilde{Q}}_l\right]\\
     &=i\sum_{k}v_{m k} u_{n k}=i\delta_{m n}.
\end{split}
\end{equation}

It confirms that the two submatrices are indeed orthonormal, as indicated by $\sum_{k}v_{m k} u_{n k}=\delta_{m n}$. Given that both $\mathcal{V}$ and $\mathcal{U}$ are unitary matrices with orthonormality $\sum_k v_{mk}v_{nk} = \sum_k u_{mk}u_{nk} = \delta_{mn}$, thus $v_{mk} = u_{mk}$, resulting in $\mathcal{V} =\mathcal{U}$. Consequently, the transformation matrix $U$ takes the form
\begin{align}
    \label{eq:U}
    U = \mathcal{U} \oplus \mathcal{U}=\begin{pmatrix}
        \mathcal{U} &0\\
        0 &\mathcal{U}
    \end{pmatrix}.
\end{align}
The submatrix $\mathcal{U}$ is solely determined by the lower right submatrix $\mathcal{H}$ in \eref{eq:bareHam_mat}. 
The vector space in bare mode and normal mode is also linked through the unitary transformation $U$. More precisely, with $\bm{\Lambda}' = U \bm{\Lambda}$, we can establish the following relationship

\begin{align}
    \label{eq:relation_xp}
    \hat{\tilde{\Phi}}_n = \sum_{m} u_{mn} \hat{\tilde{\Phi}}'_m, ~~\hat{\tilde{Q}}_n = \sum_{m} u_{mn} \hat{\tilde{Q}}'_m,
\end{align}
where $u_{mn}$ is the entry of $\mathcal{U}$. This formula expands the operators within the bare-mode representation as a linear combination of operators within the normal-mode representation.
Based on the definition of IEPR in \eref{eq:def_IEPR}, we can calculate the IEPR as follows:
\begin{equation}
\begin{split}
    r_{mn} &= \frac{\left \langle \Psi'_m \left | \frac{  \hat{\tilde{\Phi}}_n^2}{2} \right | \Psi'_m \right \rangle}
    {  \sum_l \left \langle \Psi'_m \left |  \frac{\hat{\tilde{\Phi}}_l'^2}{2} \right | \Psi'_m \right \rangle } \\
    &= \frac{ \left \langle \Psi'_m \left |   \left (\sum_k u_{kn} \hat{\tilde{\Phi}}_k' \right)^2 \right | \Psi'_m \right \rangle}
    { \sum_l \left \langle \Psi'_m \left |   \hat{\tilde{\Phi}}_l'^2 \right | \Psi'_m \right \rangle }\\
    &= \frac{  u^2_{mn}  \left \langle \Psi'_m \left |   \hat{\tilde{\Phi}}_m'^2 \right | \Psi'_m \right \rangle}
    { \left \langle \Psi'_m \left |   \hat{\tilde{\Phi}}_m'^2 \right | \Psi'_m \right \rangle } \\&= u_{mn}^2.
\end{split}
\label{eq:IEPR_der}
\end{equation}
From the above formula, we establish the relationship between IEPR and representation transformation as $u_{mn} = \pm \sqrt{r_{mn}}$. The positive and negative sign is determined by introducing a sign matrix $s_{mn}$, which can be obtained through classical electromagnetic simulation (see the main text). This relation unveils a profound physical connection between the transformation matrix and the energy participation ratio, with IEPR serving as a bridge between bare mode and normal mode.

\section{Solving the inductive energy of an element from classical electromagnetic simulation}
\label{sec:solve_IEPR}
\ In this section, we present the technical details of extracting the inductive energy of an element from classical electromagnetic simulation. First of all, the electric field $\bm{E}_m(\bm{r}) \cos(\omega_m' t)$ and eigenfrequency $\omega'_m$ corresponding to the normal mode $m$ can be obtained from classical electromagnetic simulation. Here, $\bm{E}_m(\bm{r})$ denotes the electric amplitude (peak value) at position $\bm{r}$. Subsequently, our primary objective is to compute the inductive energy $\mathcal{E}_{mn}^I$ stored in a specific element (e.g., qubit or resonator) through a postprocessing procedure.

\begin{figure*}
    \centering
    \includegraphics[width=0.9\linewidth]{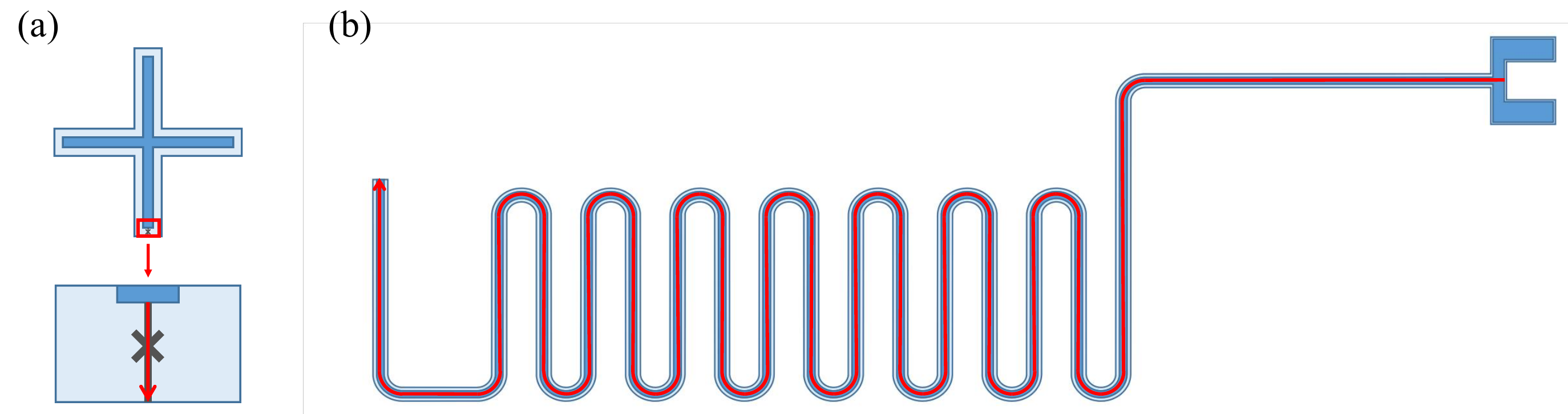}
    \caption{(a) Geometric layout of the qubit and the integral path setting (red arrow). (b) Geometric layout of the coplanar resonator and the integral path setting (red arrow).}
    \label{fig:circ_spectrum}
\end{figure*}

In electromagnetic simulations of superconducting quantum chip layouts, Josephson junction are usually modeled as a lumped inductor. It it worthwhile to note that the inductive energy associated with element containing Josephson junction comprises two components: i) the kinetic inductive energy stored in the lumped inductor, and ii) the inductive energy stored in the element's conductor. For elements not involving lumped inductor, considering only the inductive energy  stored in the element's conductor is sufficient.
Specifically, the inductive energy stored in an element is proportional to the square of the magnetic flux. Since the flux is time-dependent, the average inductive energy stored in a particular element $n$ is evaluated as half of peak inductive energy, given by
\begin{align}
    \label{eq:ind_mn}
    \mathcal{E}_{mn}^I = \frac{\Phi_{mn}^2}{4 L_n},
\end{align} 
where the peak value of flux $\Phi_{mn}$ is computed using
\begin{align}
 \label{eq:Phi_mn}
    \Phi_{mn}f_\Phi(t) = \int V_{mn}f_V(t) dt.
\end{align}
The main challenge is to compute the peak voltage $V_{mn}$ across the two ends of the $n$th element when the mode $m$ is excited and the time-dependent part $f_V(t)$. In particular, this is achieved by integrating the electric field along the path of the element from one potential node to another
\begin{align}
 \label{eq:V_mn}
    V_{mn}f_V(t) = \int_{\bm{l}_{n}}   \bm{E}_m(\bm{r})  \cdot d \bm{l} \cos(\omega_m' t)= V_{mn} \cos(\omega_m' t),
\end{align}
where $\bm{l}_{n}$ represents the integral path for component $n$.
Substituting Eqs.~\eqref{eq:Phi_mn} and \eqref{eq:V_mn} back to Eq.~\eqref{eq:ind_mn}, we arrive at the crucial result presented in the main text
\begin{equation}
        \mathcal{E}_{mn}^I = \frac{V_{mn}^2}{4L_n \omega_m'^2}.
\end{equation}
In the end, let us clarify how we determine the integral path $\bm{l}_{n}$. We distinguish between situations involving Josephson junction and those without. For elements such as qubits that contain Josephson junctions, where the kinetic inductance of the Josephson junction dominates the inductive energy, we set the integral path along the junction itself (red arrow), stretching from one end to the other, as depicted in Fig.~\ref{fig:circ_spectrum}(a). Conversely, for elements like resonators that lack Josephson junctions, the integral path is positioned along the center conductor (red arrow) as shown in Fig.~\ref{fig:circ_spectrum}(b).

\section{Reducing the many-body system to a well-handled sub-system}
\label{sec:reduce}

Large-scale superconducting quantum chips are typically described as many-body systems. However, there are instances where our focus is solely on specific elements within the system. In such cases, modeling the entire system can be prohibitively expensive. Taking for instance, the scenario discussed in \sref{subsection:linear} where we are primarily interested in the effective coupling between two qubits, even though the chip layout includes five elements. This section addresses the methodology for simplifying the many-body system into an equivalent subsystem.

As discussed before, the Hamiltonian of the many-body system described the superconducting quantum chip layout is given in \eref{eq:bareHam_mat}. Since $\hat{H}_{\rm{bare}}$ is a real symmetric matrix, it can be diagonalized into the form given in \eref{eq:normHam_mat} step by step. In particular, there exists a unitary matrix $\mathcal{A}_1 = (\mathbf{a}_{11},
        \mathbf{a}_{12},
        \cdots,
        \mathbf{a}_{1n})$ which can be employed to partially diagonalize the matrix $\mathcal{H}$, namely
\begin{equation}
    \begin{split}
    \mathcal{A}_1 \mathcal{H} \mathcal{A}_1^\dagger &= 
    \begin{pmatrix}
        \mathbf{a}_{11}^T \\
        \mathbf{a}_{12}^T\\
        \vdots\\
        \mathbf{a}_{1n}^T\\
    \end{pmatrix}
    \mathcal{H}
    \begin{pmatrix}
        \mathbf{a}_{11},
        \mathbf{a}_{12},
        \cdots,
        \mathbf{a}_{1n}
    \end{pmatrix} \\
    &=\left( \begin{array}{c|ccc}
        \omega_1'^2 &0 &\cdots &0\\
        \hline
        0 &\multicolumn{3}{c}{\multirow{3}{*}{$\mathcal{H}_{1}$}}\\
        \vdots \\
        0 
    \end{array} \right ),
\end{split}
\end{equation}
where $\mathbf{a}_{11}$ represents the eigenvector of $\mathcal{H}$ with the eigenvalue $\omega_1'^2$, while other $\mathbf{a}_{1j}~(j\in \{2, 3, \cdots, n\})$ are chosen arbitrarily to make the matrix $\mathcal{A}_1$ remains orthonormal. $\mathbf{a}_{1j}^T$ denotes the transpose operation. Regarding the nondiagonalized part $\mathcal{H}_{1}$, we can perform another unitary transformation using $\mathcal{A}_2$ to partially diagonalize it once again. By analogy, when proceeding to the $k$th step, the Hamiltonian $\mathcal{H}$ in matrix form will be expressed in terms of a $k \times k$ diagonalized submatrix and a nondiagonalized matrix of order $n-k$.
Consequently, the system Hamiltonian is written as
\begin{equation}
    \begin{split}
    \hat H_{\rm eff} &=  \sum_{m=1}^{k} \omega_m' \hat{a}_m^\dagger \hat{a}_m + \sum_{j=1}^{n-k} \omega_{j}^k \hat{a}_j^\dagger \hat{a}_j \\
    &\quad - \sum_{l\neq j} \frac{g_{lj}^k}{2} (\hat{a}_l^\dagger - \hat{a}_l) (\hat{a}_j^\dagger - \hat{a}_j).
    \end{split}
    \label{eq:ham_part_diag}
\end{equation}
In the effective Hamiltonian above, the first term represents the uncoupled subsystem and $\omega_m'$ is the normal-mode frequency. The second and third terms exactly describe the reduced $n-k$ body system, with $\omega_j^k, g_{jl}^k$ representing the effective bare frequency and effective coupling strength corrected by the other $k$ bodies. In \sref{subsection:linear}, our studied case corresponds to $n=5$ and $k=3$, where the system was reduced to an effective and simple two-body system when we explored the qubit-qubit effective coupling characteristics.

\section{CEPR techniques}
\label{Appendix:cEPR}

In the main text, we considered the frequently used scenario where only capacitive couplings are involved in the system Hamiltonian. Indeed, the IEPR method introduced in this paper can also be extended straightforwardly to the case of inductive coupling. We name it as capacitive-energy participation ratio (CEPR).
First of all,  replacing the capacitive couplings by inductive coupling, the  Hamiltonian in linear regime reads
\begin{equation}
    \hat{H}_{\rm bare}^{\rm lin} = \sum_m \frac{\hat{Q}_m^2}{2C_m} + \frac{ \hat{\Phi}_m^2}{2L_m} + \frac{1}{2}\sum_{m\neq n}\frac{M_{mn}}{L_m L_n}  \hat{\Phi}_m \hat{\Phi}_n,
\end{equation}
where the additional $M_{mn}$ denotes the mutual inductance between elements $m$ and $n$.
We make use of the same techniques applied in the main text, namely  performing operator replacements $\hat{\tilde{Q}}_m=\hat{Q}_m/\sqrt{C_m}$, $\hat{\tilde{\Phi}}_m=\hat{\Phi}_m/(\omega_m\sqrt{L_m})$ for the operator, then the Hamiltonian above is rewritten as $\hat{H}_{\rm bare}^{\rm lin}=\sum_m (\hat{\tilde{Q}}_m^2/2 + \omega_m^2\hat{\tilde{\Phi}}_m^2/2) + \sum_{m\neq n} g_{mn}\sqrt{\omega_m \omega_n}  \hat{\tilde{\Phi}}_m \hat{\tilde{\Phi}}_n$, where $g_{mn}=M_{mn} \sqrt{\omega_m \omega_n}/(2 \sqrt{L_m L_n}) $ is the coupling strength between the elements $m$ and $n$. As discussed in the main text, it can be further diagonalized: $\hat{H}_{\rm normal}^{\rm lin} = \sum_m (\hat{\tilde{Q}}_m^{\prime 2}/2 + \omega_m'^2\hat{\tilde{\Phi}}_m^{\prime 2}/2)$ through introducing an unitary transformation $U$.
Similar to the definition of IEPR, we introduce the capacitive energy participation ratio as $r^{C}_{mn} = \mathcal{E}_{mn}^C / \mathcal{E}_{m}^C$. $\mathcal E_{mn}^{C}$ represents the average capacitive energy distributed across the $n$th element, while $\mathcal E_m^{C}$ denotes the average capacitive energy  stored in the full quantum chip when the normal mode $m$ is excited. Using the same techniques, the relationship between the unitary transformation $U$ and CEPR is derived as $r^{C}_{mn} = u_{mn}^2$. To calculate $r^{C}_{mn}$ from EM simulation, the capacitive energy $\mathcal E_m^{C}$ of the full chip is the same as the electric field energy described in the main text since $\mathcal{E}_{m}^C = \mathcal{E}_m^E$. In addition, for a specific element in bare-mode representation, the capacitive energy is equal to the inductive energy, namely $\mathcal E_{mn}^{C} = \mathcal E_{mn}^{I}$. As a consequence, the calculation of CEPR is equivalent to computing IEPR. With the obtained CEPR, one can acquire the unitary matrix $U$, and further capture all of the necessary characteristic parameters and Hamiltonian as we had done in the scenario of capacitive coupling using IEPR method.

\section{Technical details of various methods}
\label{sec:simMethod}

We present a comparative analysis of the nonlinear results in \tref{tab:comp}, acquired through a variety of methods. To facilitate a comprehensive understanding, we offer technical insights into the methodologies employed.

\begin{figure*}
    \centering
    \includegraphics[width=0.86\linewidth]{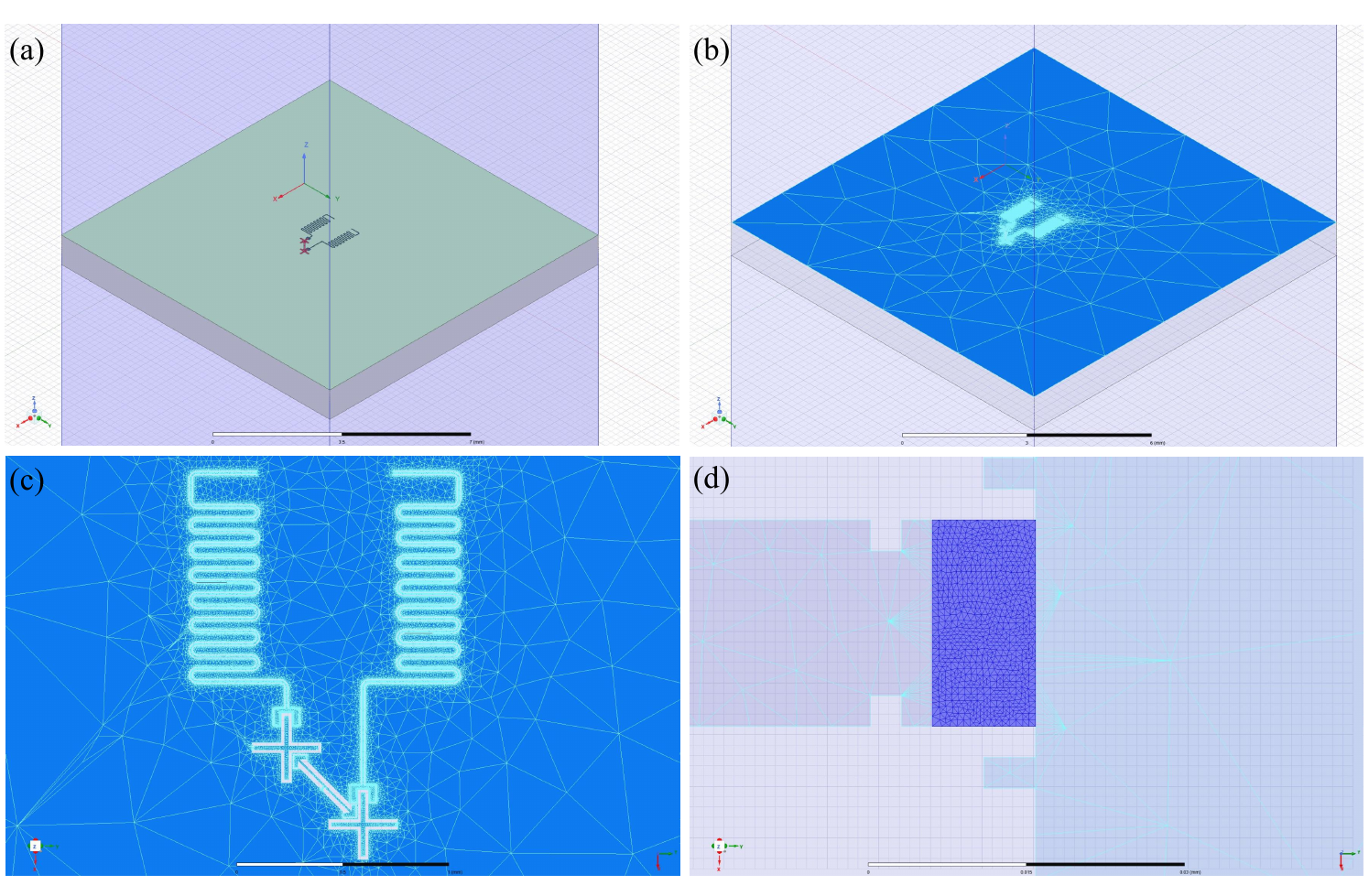}
    \caption{ANSYS HFSS model depicting the chip layout referenced in Fig. 1. (a) Presents a 3D model of the quantum chip layout. (b) Illustrates the simulation mesh configuration applied to the chip layout atop the substrate. (c) Provides an enlarged view of the core element's area, highlighting the qubit-coupler-qubit structure and the corresponding readout resonators . (d) Offers a detailed view of the finely meshed Josephson junction of $Q_2$.}
    \label{fig:simulation}
\end{figure*}

Firstly, in the context of full-wave approaches encompassing the IEPR, EPR, and BBQ methods, we adopt ANSYS HFSS software as our modeling tool. This software enables us to construct a precise 3D representation of the chip layout, vividly depicted in \fref{fig:simulation}~(a). It consists of three distinct components: the ground plane (green), qubits and couplers (red), and $1/4$ wavelength resonators (blue). Notably, the metal layer, sized at $10.5 \times 10.5$ $mm^2$, is treated as a perfect conductor. It is positioned atop a sapphire substrate (gray) with a thickness of 1 mm and a relative permittivity of $10.5$. The entire chip assembly is enclosed within a vacuum box (purple), having a height of $29$ mm.
Additionally, \fref{fig:simulation}(b) illustrates the mesh configuration employed in the final iteration of our EM simulation. Convergence criteria for the maximum frequency difference are $0.08\%$. For further clarity, \fref{fig:simulation}(c) and \fref{fig:simulation}~(d) offer close-up views of the core elements and the Josephson junction (purple) within qubit $Q_2$, respectively. These images reveal the dense meshing employed, particularly in the vicinity of the Josephson junction, which is refined compared to other regions. It is worth noting that Josephson junctions are represented as lumped inductances, effectively capturing their linear properties.

\par In the case of the IEPR and EPR methods, both of which rely on eigenmode simulations, our initial output typically comprises eigenmode frequencies and the associated electric field distributions of normal modes. However, it is paramount to underscore the criticality of selecting the appropriate mode for subsequent calculations. As expounded upon in \sref{sec:IEPR}, a pivotal step in this process involves the identification of the fundamental normal mode. Armed with the knowledge of these normal modes, both the IEPR and EPR methods facilitate the efficient derivation of the corresponding characteristic parameters. At the crux of the IEPR method procedure, as depicted in \fref{fig:workflow}, lies the pivotal utilization of postprocessing techniques. These techniques serve a dual role, enabling the resolution not only of the linear bare frequencies of the qubit and coupler, as well as their coupling strength, but also the determination of all the nonlinear characteristic parameters.

Alternatively, the nonlinear results in normal-mode representation obtained with IEPR method can be cross checked with the aid of numerical methodology. In particular, the numerical diagonalization is undertaken to derive the eigenenergy spectrum of the bare-mode Hamiltonian $\hat{H}^{\rm nl}_{\rm bare}$. $E_{ij}$ denotes the corresponding eigenenergy of the eigenstate $\vert ij \rangle$ ($i,j=0,1,2$, $\cdots$).
Using the spectrum information, we are able to discern the fundamental modes associated with the renormalization frequencies of the qubit and coupler. In the regime of interest, these identified modes served as the foundation for applying the following relationships
\begin{align}
    &\omega_q^{\rm nl \prime} = E_{01}\\
    &\omega_c^{\rm nl \prime} = E_{10}\\
    &\alpha_q^{\prime} = E_{02} - 2E_{01}\\
    &\alpha_c^{\prime} = E_{20} - 2E_{10}\\
    &\chi_{qc} = E_{11} - E_{01} - E_{10}
\end{align}
to compute the relevant nonlinear parameters. Note that we set $E_{00}=0$. Of note, this methodology lends itself to extension for more intricate many-body systems. By comparing the bare frequencies, we can ascertain the renormalized fundamental mode pertaining to the respective element. This expanded framework allows us to correlate quantum states at higher energy levels and subsequently calculate the cross-Kerr and self-Kerr parameters based on the energy spectrum.

Subsequently, we transition to the EPR method. Utilizing the identical EM simulation outcomes as IEPR approach, we proceed to calculate the energy participation ratios, denoted as $p_{qq}, p_{cq}$ for the qubit and coupler modes. These calculations, along with the knowledge of the Josephson energies, facilitated the derivation of the nonlinear parameters \cite{minev2021energy}.
\begin{align}
    &\alpha_q^{\prime} = -\frac{\omega_q^{\prime 2} p_{qq}^2}{8E_J^q} -\frac{\omega_q^{\prime 2}p_{qc}^2}{8E_J^c},\\
    &\alpha_q^{\prime} = -\frac{\omega_c^{\prime 2} p_{cc}^2}{8E_J^c} - \frac{\omega_c^{\prime 2}p_{cq}^2}{8E_J^q},\\
    &\chi_{qc} = -\frac{\omega_q \omega_c p_{qq}p_{cq}}{4E_J^q} - \frac{\omega_q \omega_c p_{qc}p_{cc}}{4E_J^c},\\
    &\omega_q^{\prime \rm nl} = \omega_q^\prime + \alpha_q^{\prime} + \frac{1}{2}\chi_{qc},\\
    &\omega_c^{\prime \rm nl} = \omega_c^\prime + \alpha_c^{\prime} + \frac{1}{2}\chi_{qc}.
\end{align}
It is noteworthy that EPR benefits from a well-established Python package, pyEPR \cite{pyEPR}, which we utilized for the simulations, harnessing the capabilities of this robust tool.

Regarding the BBQ method, as elucidated in Ref.~\cite{nigg2012black}, our approach involved configuring corresponding ports at both ends of the Josephson junction within the 3D model, as depicted in \fref{fig:simulation}. Subsequently, through EM simulation using HFSS, we obtained the admittance parameters (Y parameters) and their derivatives for the port associated with the normal mode $\omega_p$ of the element $p$. This information allowed us to calculate the effective capacitance and effective inductance. Ultimately, employing Eq.~(7) of Ref.~\cite{nigg2012black}, we resolved the nonlinear parameters.
\begin{figure}[b]
    \centering
    \includegraphics[width=\linewidth]{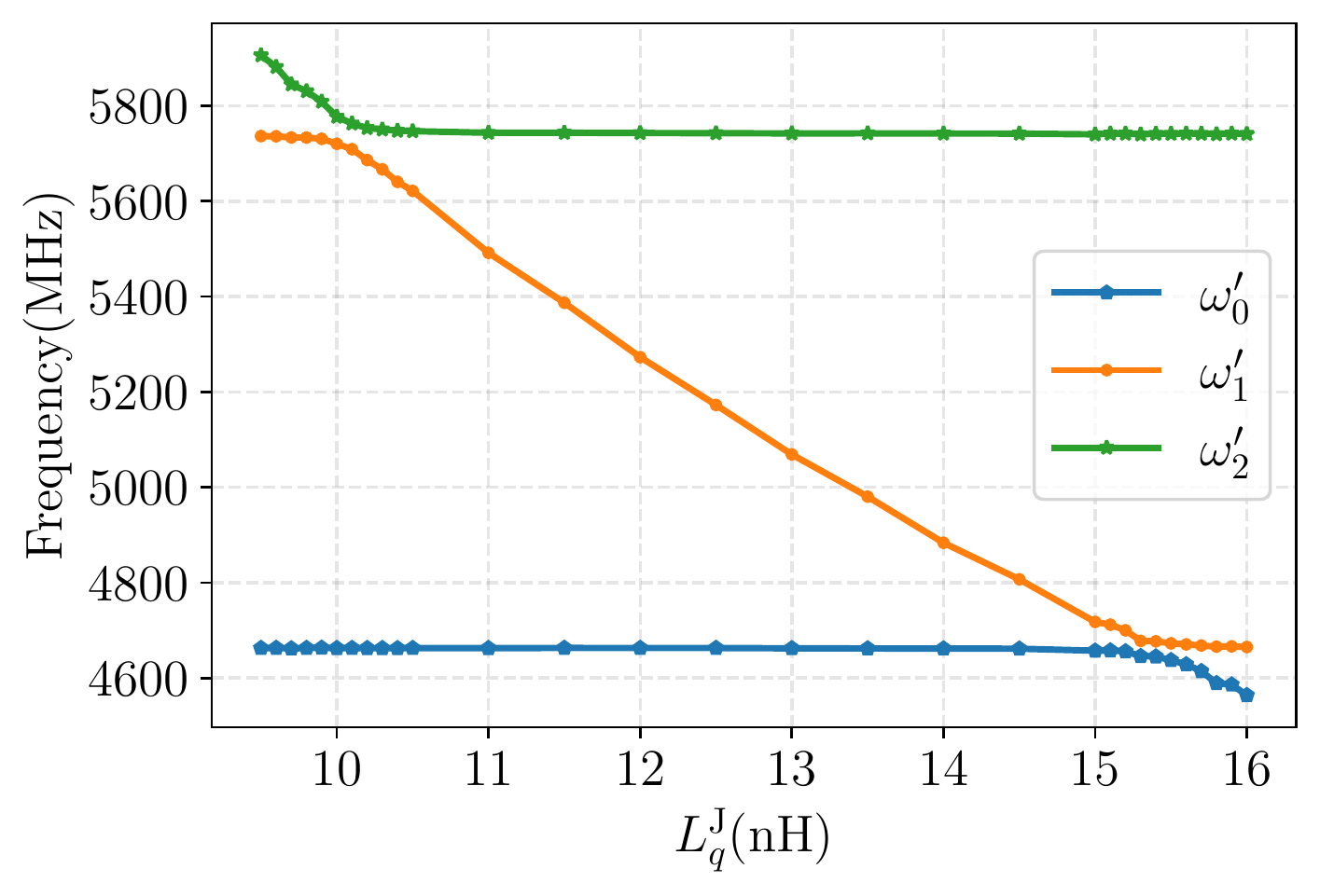}
    \caption{The spectrum of the system depicted in Fig.~\ref{fig:transmon_3Dcavity}, solved using NMS method. Three normal modes exhibit variation in response to changes in the inductance of Josephson junction. Two anticrossing points appear at 10.1 nH and 15.4 nH.}
    \label{fig:NMS_3Dcavity}
\end{figure}
\begin{figure*}[t]
    \centering
    \includegraphics[width=\linewidth]{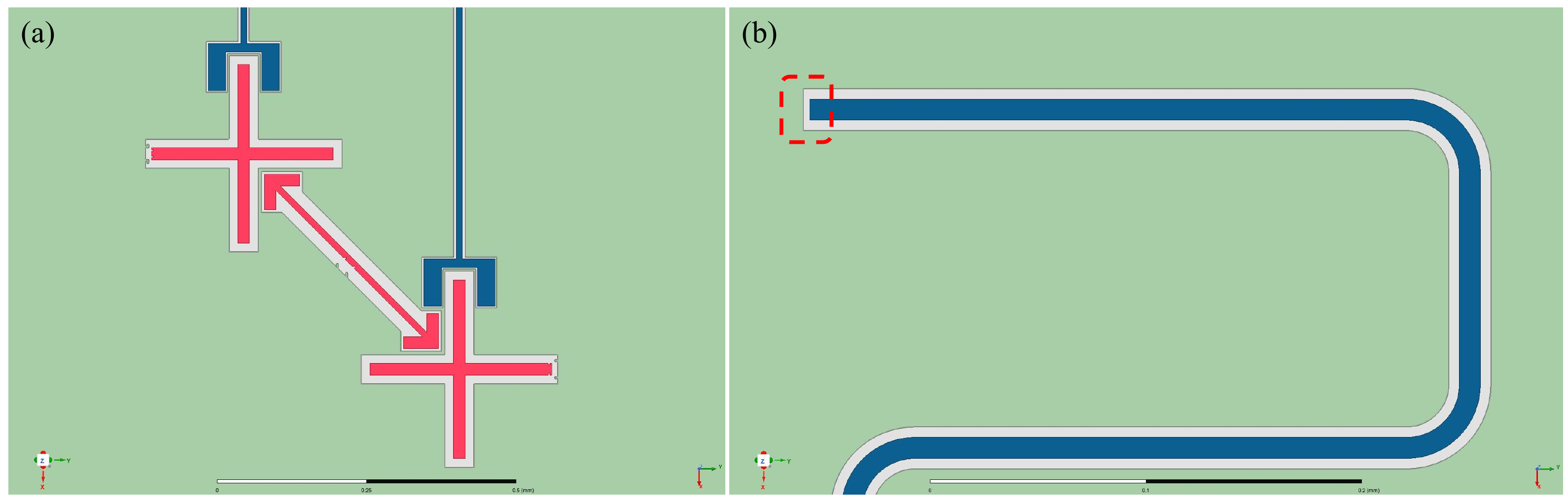}
    \caption{ANSYS Q3D model of the chip layout. (a) Depicts the qubit-coupler-qubit structure with the Josephson junction omitted. (b) Illustrates the resonator configuration, where the grounded end is marginally truncated (red dotted box) to avoid direct contact with the ground.}
    \label{fig:q3d}
\end{figure*}

Within the realm of full-wave simulation, an alternative approach known as the normal-mode Simulation (NMS) method, as detailed in Ref.~\cite{dubyna2020inter, yanay2022mediated}, stands apart from other methodologies. Notably, the NMS method exhibits certain limitations--it is primarily applicable to two-body systems and excels in accurately resolving the bare-mode parameters exclusively under conditions of full resonance. In nonresonant scenarios, the determination of coupling strength is approached with an approximation.
To elucidate further, consider a resonance case involving two identical qubits, where their bare frequencies are equivalent, i.e., $\omega_1 = \omega_2$. The coupling strength between these qubits is denoted as $g_{12}$. The interplay between the diagonalized normal-mode frequencies, $\omega_1^\prime$ and $\omega_2^\prime$, of the system and the bare state frequencies, along with the coupling strength, can be described as follows:
\begin{align}
    &\omega_1 = \omega_2 = \frac{\omega_1^\prime + \omega_2^\prime}{2}, \\
    \label{eq:g_NMS}
    &g_{12} = \left | \frac{\omega_1^\prime - \omega_2^\prime}{2} \right |.
\end{align}
Nonetheless, the NMS method encounters challenges when applied to nonresonant cases, requiring a strategic approach. To address this, a workaround is employed. By capitalizing on the relationship between coupling strength and bare frequency, expressed as $g \propto \sqrt{\omega_1 \omega_2}$, we manipulate the inductance value of the Josephson junction. This manipulation serves to adjust the qubit's frequency to align with resonance conditions, facilitating the determination of coupling strength for resonance through \eref{eq:g_NMS}. Subsequently, the resonant frequency is substituted with the actual frequency to ascertain the coupling strength in nonresonant scenarios. 
To illustrate this approach, consider the example of the double 3D cavities coupled with a transmon qubit depicted in Fig.~\ref{fig:transmon_3Dcavity}. Here, the inductance value of the qubit is adjusted to find the resonance point, as depicted in \fref{fig:NMS_3Dcavity}, which illustrates the three frequencies of different normal modes. The resonance point, i.e., the minimum frequency difference, is determined at 10.1 and 15.4 nH. In particular, 10.1 nH corresponds to the resonance point of ``Qubit" and ``Cavity2", where the coupling strength is 27.02 MHz and the bare frequency is 5736.08 MHz. Replacing the frequency with 5068.88 and 5742.23 MHz at 13 nH, the actual coupling strength is evaluated as 25.41 MHz. Similarly, 15.4 nH corresponds to the resonance point of ``Qubit" and ``cavity1", where the coupling strength is 15.99 MHz and the bare frequency is 4660.97 MHz. We replace the frequencies by 4662.27 and 5068.88 MHz at 13 nH, obtaining the actual coupling strength as 16.67 MHz. These results align with counterpart obtained using the IEPR method, as presented in \tref{tab:transmon_3Dcavity}. 
It is worthwhile to point out in the quest to find the resonance point, the NMS method necessitates multiple simulation runs to sweep through different frequencies, whereas the IEPR method requires only a single simulation at the actual frequency, demonstrating its superior efficiency.

Finally, we delve into the relatively expedient LC method, grounded in capacitance simulation of static fields. In contrast to full-wave methodologies, this approach does not involve high-frequency electromagnetic field oscillations, resulting in reduced simulation complexity. In this regard, we employ ANSYS Q3D for obtaining the capacitance matrix. This involves treating continuous conductors as equipotential entities and modeling the chip layout as an equivalent circuit. The simulation centers on evaluating the capacitance between various conductors and the known linear inductance values of Josephson junctions.
It is crucial to emphasize that in capacitance simulation, depicted in \fref{fig:q3d}(a), the Josephson junction is excluded, while the size and boundary conditions of the remaining structure remain consistent. Additionally, since one end of the $1/4$ wavelength resonator is grounded, as indicated by the red dotted box in \fref{fig:q3d}(b), a small segment has to be removed from the grounded end, with frequency corrections (using the relation between $\lambda/4$ and $\lambda/2$ resonators) applied subsequently.
Following the principles of the superconducting quantum circuit quantization theory \cite{blais2021circuit}, we can ascertain the bare frequency, coupling strength, and anharmonicity of the individual elements. Subsequently, employing analogous diagonalization steps introduced before, one can further solve for renormalized frequency, self-Kerr and cross-Kerr parameters.

\bibliography{reference}

\begin{thebibliography}{45}%
\makeatletter
\providecommand \@ifxundefined [1]{%
 \@ifx{#1\undefined}
}%
\providecommand \@ifnum [1]{%
 \ifnum #1\expandafter \@firstoftwo
 \else \expandafter \@secondoftwo
 \fi
}%
\providecommand \@ifx [1]{%
 \ifx #1\expandafter \@firstoftwo
 \else \expandafter \@secondoftwo
 \fi
}%
\providecommand \natexlab [1]{#1}%
\providecommand \enquote  [1]{``#1''}%
\providecommand \bibnamefont  [1]{#1}%
\providecommand \bibfnamefont [1]{#1}%
\providecommand \citenamefont [1]{#1}%
\providecommand \href@noop [0]{\@secondoftwo}%
\providecommand \href [0]{\begingroup \@sanitize@url \@href}%
\providecommand \@href[1]{\@@startlink{#1}\@@href}%
\providecommand \@@href[1]{\endgroup#1\@@endlink}%
\providecommand \@sanitize@url [0]{\catcode `\\12\catcode `\$12\catcode `\&12\catcode `\#12\catcode `\^12\catcode `\_12\catcode `\%12\relax}%
\providecommand \@@startlink[1]{}%
\providecommand \@@endlink[0]{}%
\providecommand \url  [0]{\begingroup\@sanitize@url \@url }%
\providecommand \@url [1]{\endgroup\@href {#1}{\urlprefix }}%
\providecommand \urlprefix  [0]{URL }%
\providecommand \Eprint [0]{\href }%
\providecommand \doibase [0]{https://doi.org/}%
\providecommand \selectlanguage [0]{\@gobble}%
\providecommand \bibinfo  [0]{\@secondoftwo}%
\providecommand \bibfield  [0]{\@secondoftwo}%
\providecommand \translation [1]{[#1]}%
\providecommand \BibitemOpen [0]{}%
\providecommand \bibitemStop [0]{}%
\providecommand \bibitemNoStop [0]{.\EOS\space}%
\providecommand \EOS [0]{\spacefactor3000\relax}%
\providecommand \BibitemShut  [1]{\csname bibitem#1\endcsname}%
\let\auto@bib@innerbib\@empty
\bibitem [{\citenamefont {Arute}\ \emph {et~al.}(2019)\citenamefont {Arute}, \citenamefont {Arya}, \citenamefont {Babbush}, \citenamefont {Bacon}, \citenamefont {Bardin}, \citenamefont {Barends}, \citenamefont {Biswas}, \citenamefont {Boixo}, \citenamefont {Brandao}, \citenamefont {Buell} \emph {et~al.}}]{arute2019quantum}%
  \BibitemOpen
  \bibfield  {author} {\bibinfo {author} {\bibfnamefont {F.}~\bibnamefont {Arute}}, \bibinfo {author} {\bibfnamefont {K.}~\bibnamefont {Arya}}, \bibinfo {author} {\bibfnamefont {R.}~\bibnamefont {Babbush}}, \bibinfo {author} {\bibfnamefont {D.}~\bibnamefont {Bacon}}, \bibinfo {author} {\bibfnamefont {J.~C.}\ \bibnamefont {Bardin}}, \bibinfo {author} {\bibfnamefont {R.}~\bibnamefont {Barends}}, \bibinfo {author} {\bibfnamefont {R.}~\bibnamefont {Biswas}}, \bibinfo {author} {\bibfnamefont {S.}~\bibnamefont {Boixo}}, \bibinfo {author} {\bibfnamefont {F.~G.}\ \bibnamefont {Brandao}}, \bibinfo {author} {\bibfnamefont {D.~A.}\ \bibnamefont {Buell}}, \emph {et~al.},\ }\bibfield  {title} {\bibinfo {title} {Quantum supremacy using a programmable superconducting processor},\ }\href {https://doi.org/https://doi.org/10.1038/s41586-019-1666-5} {\bibfield  {journal} {\bibinfo  {journal} {Nature}\ }\textbf {\bibinfo {volume} {574}},\ \bibinfo {pages} {505} (\bibinfo {year} {2019})}\BibitemShut {NoStop}%
\bibitem [{\citenamefont {Gong}\ \emph {et~al.}(2021)\citenamefont {Gong}, \citenamefont {Wang}, \citenamefont {Zha}, \citenamefont {Chen}, \citenamefont {Huang}, \citenamefont {Wu}, \citenamefont {Zhu}, \citenamefont {Zhao}, \citenamefont {Li}, \citenamefont {Guo} \emph {et~al.}}]{gong2021quantum}%
  \BibitemOpen
  \bibfield  {author} {\bibinfo {author} {\bibfnamefont {M.}~\bibnamefont {Gong}}, \bibinfo {author} {\bibfnamefont {S.}~\bibnamefont {Wang}}, \bibinfo {author} {\bibfnamefont {C.}~\bibnamefont {Zha}}, \bibinfo {author} {\bibfnamefont {M.-C.}\ \bibnamefont {Chen}}, \bibinfo {author} {\bibfnamefont {H.-L.}\ \bibnamefont {Huang}}, \bibinfo {author} {\bibfnamefont {Y.}~\bibnamefont {Wu}}, \bibinfo {author} {\bibfnamefont {Q.}~\bibnamefont {Zhu}}, \bibinfo {author} {\bibfnamefont {Y.}~\bibnamefont {Zhao}}, \bibinfo {author} {\bibfnamefont {S.}~\bibnamefont {Li}}, \bibinfo {author} {\bibfnamefont {S.}~\bibnamefont {Guo}}, \emph {et~al.},\ }\bibfield  {title} {\bibinfo {title} {Quantum walks on a programmable two-dimensional 62-qubit superconducting processor},\ }\href {https://doi.org/10.1126/science.abg7812} {\bibfield  {journal} {\bibinfo  {journal} {Science}\ }\textbf {\bibinfo {volume} {372}},\ \bibinfo {pages} {948} (\bibinfo {year} {2021})}\BibitemShut {NoStop}%
\bibitem [{\citenamefont {Wu}\ \emph {et~al.}(2021)\citenamefont {Wu}, \citenamefont {Bao}, \citenamefont {Cao}, \citenamefont {Chen}, \citenamefont {Chen}, \citenamefont {Chen}, \citenamefont {Chung}, \citenamefont {Deng}, \citenamefont {Du}, \citenamefont {Fan} \emph {et~al.}}]{wu2021strong}%
  \BibitemOpen
  \bibfield  {author} {\bibinfo {author} {\bibfnamefont {Y.}~\bibnamefont {Wu}}, \bibinfo {author} {\bibfnamefont {W.-S.}\ \bibnamefont {Bao}}, \bibinfo {author} {\bibfnamefont {S.}~\bibnamefont {Cao}}, \bibinfo {author} {\bibfnamefont {F.}~\bibnamefont {Chen}}, \bibinfo {author} {\bibfnamefont {M.-C.}\ \bibnamefont {Chen}}, \bibinfo {author} {\bibfnamefont {X.}~\bibnamefont {Chen}}, \bibinfo {author} {\bibfnamefont {T.-H.}\ \bibnamefont {Chung}}, \bibinfo {author} {\bibfnamefont {H.}~\bibnamefont {Deng}}, \bibinfo {author} {\bibfnamefont {Y.}~\bibnamefont {Du}}, \bibinfo {author} {\bibfnamefont {D.}~\bibnamefont {Fan}}, \emph {et~al.},\ }\bibfield  {title} {\bibinfo {title} {Strong quantum computational advantage using a superconducting quantum processor},\ }\href {https://doi.org/https://doi.org/10.1103/PhysRevLett.127.180501} {\bibfield  {journal} {\bibinfo  {journal} {Physical Review Letters}\ }\textbf {\bibinfo {volume} {127}},\ \bibinfo {pages} {180501} (\bibinfo {year} {2021})}\BibitemShut {NoStop}%
\bibitem [{\citenamefont {Chow}\ \emph {et~al.}(2021)\citenamefont {Chow}, \citenamefont {Dial},\ and\ \citenamefont {Gambetta}}]{chow2021ibm}%
  \BibitemOpen
  \bibfield  {author} {\bibinfo {author} {\bibfnamefont {J.}~\bibnamefont {Chow}}, \bibinfo {author} {\bibfnamefont {O.}~\bibnamefont {Dial}},\ and\ \bibinfo {author} {\bibfnamefont {J.}~\bibnamefont {Gambetta}},\ }\bibfield  {title} {\bibinfo {title} {{IBM} quantum breaks the 100-qubit processor barrier},\ }\href {https://research.ibm.com/blog/127-qubit-quantum-processor-eagle} {\bibfield  {journal} {\bibinfo  {journal} {IBM Research Blog}\ } (\bibinfo {year} {2021})}\BibitemShut {NoStop}%
\bibitem [{\citenamefont {Rajeev}\ \emph {et~al.}(2023)\citenamefont {Rajeev}, \citenamefont {Igor}, \citenamefont {Richard}, \citenamefont {Trond}, \citenamefont {Markus}, \citenamefont {Frank}, \citenamefont {Kunal}, \citenamefont {Abraham}, \citenamefont {Juan}, \citenamefont {Ryan} \emph {et~al.}}]{google2023suppressing}%
  \BibitemOpen
  \bibfield  {author} {\bibinfo {author} {\bibfnamefont {A.}~\bibnamefont {Rajeev}}, \bibinfo {author} {\bibfnamefont {A.}~\bibnamefont {Igor}}, \bibinfo {author} {\bibfnamefont {A.}~\bibnamefont {Richard}}, \bibinfo {author} {\bibfnamefont {I.~A.}\ \bibnamefont {Trond}}, \bibinfo {author} {\bibfnamefont {A.}~\bibnamefont {Markus}}, \bibinfo {author} {\bibfnamefont {A.}~\bibnamefont {Frank}}, \bibinfo {author} {\bibfnamefont {A.}~\bibnamefont {Kunal}}, \bibinfo {author} {\bibfnamefont {A.}~\bibnamefont {Abraham}}, \bibinfo {author} {\bibfnamefont {A.}~\bibnamefont {Juan}}, \bibinfo {author} {\bibfnamefont {B.}~\bibnamefont {Ryan}}, \emph {et~al.},\ }\bibfield  {title} {\bibinfo {title} {Suppressing quantum errors by scaling a surface code logical qubit},\ }\href {https://doi.org/10.1038/s41586-022-05434-1} {\bibfield  {journal} {\bibinfo  {journal} {Nature}\ }\textbf {\bibinfo {volume} {614}},\ \bibinfo {pages} {676} (\bibinfo {year} {2023})}\BibitemShut {NoStop}%
\bibitem [{\citenamefont {Alt}(2022)}]{alt2022potentials}%
  \BibitemOpen
  \bibfield  {author} {\bibinfo {author} {\bibfnamefont {R.}~\bibnamefont {Alt}},\ }\bibfield  {title} {\bibinfo {title} {On the potentials of quantum computing--an interview with heike riel from ibm research},\ }\href {https://doi.org/10.1007/s12525-022-00616-1} {\bibfield  {journal} {\bibinfo  {journal} {Electronic Markets}\ }\textbf {\bibinfo {volume} {32}},\ \bibinfo {pages} {2537} (\bibinfo {year} {2022})}\BibitemShut {NoStop}%
\bibitem [{\citenamefont {Rigetti}\ \emph {et~al.}(2012)\citenamefont {Rigetti}, \citenamefont {Gambetta}, \citenamefont {Poletto}, \citenamefont {Plourde}, \citenamefont {Chow}, \citenamefont {C{\'o}rcoles}, \citenamefont {Smolin}, \citenamefont {Merkel}, \citenamefont {Rozen}, \citenamefont {Keefe} \emph {et~al.}}]{rigetti2012superconducting}%
  \BibitemOpen
  \bibfield  {author} {\bibinfo {author} {\bibfnamefont {C.}~\bibnamefont {Rigetti}}, \bibinfo {author} {\bibfnamefont {J.~M.}\ \bibnamefont {Gambetta}}, \bibinfo {author} {\bibfnamefont {S.}~\bibnamefont {Poletto}}, \bibinfo {author} {\bibfnamefont {B.~L.}\ \bibnamefont {Plourde}}, \bibinfo {author} {\bibfnamefont {J.~M.}\ \bibnamefont {Chow}}, \bibinfo {author} {\bibfnamefont {A.~D.}\ \bibnamefont {C{\'o}rcoles}}, \bibinfo {author} {\bibfnamefont {J.~A.}\ \bibnamefont {Smolin}}, \bibinfo {author} {\bibfnamefont {S.~T.}\ \bibnamefont {Merkel}}, \bibinfo {author} {\bibfnamefont {J.~R.}\ \bibnamefont {Rozen}}, \bibinfo {author} {\bibfnamefont {G.~A.}\ \bibnamefont {Keefe}}, \emph {et~al.},\ }\bibfield  {title} {\bibinfo {title} {Superconducting qubit in a waveguide cavity with a coherence time approaching 0.1 ms},\ }\href {https://doi.org/10.1103/PhysRevB.86.100506} {\bibfield  {journal} {\bibinfo  {journal} {Physical Review B}\ }\textbf {\bibinfo {volume} {86}},\ \bibinfo {pages} {100506} (\bibinfo {year}
  {2012})}\BibitemShut {NoStop}%
\bibitem [{\citenamefont {Nguyen}\ \emph {et~al.}(2019)\citenamefont {Nguyen}, \citenamefont {Lin}, \citenamefont {Somoroff}, \citenamefont {Mencia}, \citenamefont {Grabon},\ and\ \citenamefont {Manucharyan}}]{nguyen2019high}%
  \BibitemOpen
  \bibfield  {author} {\bibinfo {author} {\bibfnamefont {L.~B.}\ \bibnamefont {Nguyen}}, \bibinfo {author} {\bibfnamefont {Y.-H.}\ \bibnamefont {Lin}}, \bibinfo {author} {\bibfnamefont {A.}~\bibnamefont {Somoroff}}, \bibinfo {author} {\bibfnamefont {R.}~\bibnamefont {Mencia}}, \bibinfo {author} {\bibfnamefont {N.}~\bibnamefont {Grabon}},\ and\ \bibinfo {author} {\bibfnamefont {V.~E.}\ \bibnamefont {Manucharyan}},\ }\bibfield  {title} {\bibinfo {title} {High-coherence fluxonium qubit},\ }\href {https://doi.org/10.1103/PhysRevX.9.041041} {\bibfield  {journal} {\bibinfo  {journal} {Physical Review X}\ }\textbf {\bibinfo {volume} {9}},\ \bibinfo {pages} {041041} (\bibinfo {year} {2019})}\BibitemShut {NoStop}%
\bibitem [{\citenamefont {Place}\ \emph {et~al.}(2021)\citenamefont {Place}, \citenamefont {Rodgers}, \citenamefont {Mundada}, \citenamefont {Smitham}, \citenamefont {Fitzpatrick}, \citenamefont {Leng}, \citenamefont {Premkumar}, \citenamefont {Bryon}, \citenamefont {Vrajitoarea}, \citenamefont {Sussman} \emph {et~al.}}]{place2021new}%
  \BibitemOpen
  \bibfield  {author} {\bibinfo {author} {\bibfnamefont {A.~P.}\ \bibnamefont {Place}}, \bibinfo {author} {\bibfnamefont {L.~V.}\ \bibnamefont {Rodgers}}, \bibinfo {author} {\bibfnamefont {P.}~\bibnamefont {Mundada}}, \bibinfo {author} {\bibfnamefont {B.~M.}\ \bibnamefont {Smitham}}, \bibinfo {author} {\bibfnamefont {M.}~\bibnamefont {Fitzpatrick}}, \bibinfo {author} {\bibfnamefont {Z.}~\bibnamefont {Leng}}, \bibinfo {author} {\bibfnamefont {A.}~\bibnamefont {Premkumar}}, \bibinfo {author} {\bibfnamefont {J.}~\bibnamefont {Bryon}}, \bibinfo {author} {\bibfnamefont {A.}~\bibnamefont {Vrajitoarea}}, \bibinfo {author} {\bibfnamefont {S.}~\bibnamefont {Sussman}}, \emph {et~al.},\ }\bibfield  {title} {\bibinfo {title} {New material platform for superconducting transmon qubits with coherence times exceeding 0.3 milliseconds},\ }\href {https://doi.org/10.1038/s41467-021-22030-5} {\bibfield  {journal} {\bibinfo  {journal} {Nature Communications}\ }\textbf {\bibinfo {volume} {12}},\ \bibinfo {pages} {1779} (\bibinfo
  {year} {2021})}\BibitemShut {NoStop}%
\bibitem [{\citenamefont {Wang}\ \emph {et~al.}(2022)\citenamefont {Wang}, \citenamefont {Li}, \citenamefont {Xu}, \citenamefont {Li}, \citenamefont {Wang}, \citenamefont {Yang}, \citenamefont {Mi}, \citenamefont {Liang}, \citenamefont {Su}, \citenamefont {Yang} \emph {et~al.}}]{wang2022towards}%
  \BibitemOpen
  \bibfield  {author} {\bibinfo {author} {\bibfnamefont {C.}~\bibnamefont {Wang}}, \bibinfo {author} {\bibfnamefont {X.}~\bibnamefont {Li}}, \bibinfo {author} {\bibfnamefont {H.}~\bibnamefont {Xu}}, \bibinfo {author} {\bibfnamefont {Z.}~\bibnamefont {Li}}, \bibinfo {author} {\bibfnamefont {J.}~\bibnamefont {Wang}}, \bibinfo {author} {\bibfnamefont {Z.}~\bibnamefont {Yang}}, \bibinfo {author} {\bibfnamefont {Z.}~\bibnamefont {Mi}}, \bibinfo {author} {\bibfnamefont {X.}~\bibnamefont {Liang}}, \bibinfo {author} {\bibfnamefont {T.}~\bibnamefont {Su}}, \bibinfo {author} {\bibfnamefont {C.}~\bibnamefont {Yang}}, \emph {et~al.},\ }\bibfield  {title} {\bibinfo {title} {Towards practical quantum computers: Transmon qubit with a lifetime approaching 0.5 milliseconds},\ }\href {https://doi.org/10.1038/s41534-021-00510-2} {\bibfield  {journal} {\bibinfo  {journal} {npj Quantum Information}\ }\textbf {\bibinfo {volume} {8}},\ \bibinfo {pages} {3} (\bibinfo {year} {2022})}\BibitemShut {NoStop}%
\bibitem [{\citenamefont {McArdle}\ \emph {et~al.}(2019)\citenamefont {McArdle}, \citenamefont {Jones}, \citenamefont {Endo}, \citenamefont {Li}, \citenamefont {Benjamin},\ and\ \citenamefont {Yuan}}]{mcardle2019variational}%
  \BibitemOpen
  \bibfield  {author} {\bibinfo {author} {\bibfnamefont {S.}~\bibnamefont {McArdle}}, \bibinfo {author} {\bibfnamefont {T.}~\bibnamefont {Jones}}, \bibinfo {author} {\bibfnamefont {S.}~\bibnamefont {Endo}}, \bibinfo {author} {\bibfnamefont {Y.}~\bibnamefont {Li}}, \bibinfo {author} {\bibfnamefont {S.~C.}\ \bibnamefont {Benjamin}},\ and\ \bibinfo {author} {\bibfnamefont {X.}~\bibnamefont {Yuan}},\ }\bibfield  {title} {\bibinfo {title} {Variational ansatz-based quantum simulation of imaginary time evolution},\ }\href {https://doi.org/10.1038/s41534-019-0187-2} {\bibfield  {journal} {\bibinfo  {journal} {npj Quantum Information}\ }\textbf {\bibinfo {volume} {5}},\ \bibinfo {pages} {75} (\bibinfo {year} {2019})}\BibitemShut {NoStop}%
\bibitem [{\citenamefont {Arute}\ \emph {et~al.}(2020)\citenamefont {Arute}, \citenamefont {Arya}, \citenamefont {Babbush}, \citenamefont {Bacon}, \citenamefont {Bardin}, \citenamefont {Barends}, \citenamefont {Boixo}, \citenamefont {Broughton}, \citenamefont {Buckley} \emph {et~al.}}]{google2020hartree}%
  \BibitemOpen
  \bibfield  {author} {\bibinfo {author} {\bibfnamefont {F.}~\bibnamefont {Arute}}, \bibinfo {author} {\bibfnamefont {K.}~\bibnamefont {Arya}}, \bibinfo {author} {\bibfnamefont {R.}~\bibnamefont {Babbush}}, \bibinfo {author} {\bibfnamefont {D.}~\bibnamefont {Bacon}}, \bibinfo {author} {\bibfnamefont {J.~C.}\ \bibnamefont {Bardin}}, \bibinfo {author} {\bibfnamefont {R.}~\bibnamefont {Barends}}, \bibinfo {author} {\bibfnamefont {S.}~\bibnamefont {Boixo}}, \bibinfo {author} {\bibfnamefont {M.}~\bibnamefont {Broughton}}, \bibinfo {author} {\bibfnamefont {B.~B.}\ \bibnamefont {Buckley}}, \emph {et~al.},\ }\bibfield  {title} {\bibinfo {title} {Hartree-fock on a superconducting qubit quantum computer},\ }\href {https://doi.org/10.1126/science.abb9811} {\bibfield  {journal} {\bibinfo  {journal} {Science}\ }\textbf {\bibinfo {volume} {369}},\ \bibinfo {pages} {1084} (\bibinfo {year} {2020})}\BibitemShut {NoStop}%
\bibitem [{\citenamefont {Eddins}\ \emph {et~al.}(2022)\citenamefont {Eddins}, \citenamefont {Motta}, \citenamefont {Gujarati}, \citenamefont {Bravyi}, \citenamefont {Mezzacapo}, \citenamefont {Hadfield},\ and\ \citenamefont {Sheldon}}]{eddins2022doubling}%
  \BibitemOpen
  \bibfield  {author} {\bibinfo {author} {\bibfnamefont {A.}~\bibnamefont {Eddins}}, \bibinfo {author} {\bibfnamefont {M.}~\bibnamefont {Motta}}, \bibinfo {author} {\bibfnamefont {T.~P.}\ \bibnamefont {Gujarati}}, \bibinfo {author} {\bibfnamefont {S.}~\bibnamefont {Bravyi}}, \bibinfo {author} {\bibfnamefont {A.}~\bibnamefont {Mezzacapo}}, \bibinfo {author} {\bibfnamefont {C.}~\bibnamefont {Hadfield}},\ and\ \bibinfo {author} {\bibfnamefont {S.}~\bibnamefont {Sheldon}},\ }\bibfield  {title} {\bibinfo {title} {Doubling the size of quantum simulators by entanglement forging},\ }\href {https://doi.org/10.1103/PRXQuantum.3.010309} {\bibfield  {journal} {\bibinfo  {journal} {PRX Quantum}\ }\textbf {\bibinfo {volume} {3}},\ \bibinfo {pages} {010309} (\bibinfo {year} {2022})}\BibitemShut {NoStop}%
\bibitem [{\citenamefont {Mi}\ \emph {et~al.}(2022)\citenamefont {Mi}, \citenamefont {Ippoliti}, \citenamefont {Quintana}, \citenamefont {Greene}, \citenamefont {Chen}, \citenamefont {Gross}, \citenamefont {Arute}, \citenamefont {Arya}, \citenamefont {Atalaya}, \citenamefont {Babbush} \emph {et~al.}}]{mi2022time}%
  \BibitemOpen
  \bibfield  {author} {\bibinfo {author} {\bibfnamefont {X.}~\bibnamefont {Mi}}, \bibinfo {author} {\bibfnamefont {M.}~\bibnamefont {Ippoliti}}, \bibinfo {author} {\bibfnamefont {C.}~\bibnamefont {Quintana}}, \bibinfo {author} {\bibfnamefont {A.}~\bibnamefont {Greene}}, \bibinfo {author} {\bibfnamefont {Z.}~\bibnamefont {Chen}}, \bibinfo {author} {\bibfnamefont {J.}~\bibnamefont {Gross}}, \bibinfo {author} {\bibfnamefont {F.}~\bibnamefont {Arute}}, \bibinfo {author} {\bibfnamefont {K.}~\bibnamefont {Arya}}, \bibinfo {author} {\bibfnamefont {J.}~\bibnamefont {Atalaya}}, \bibinfo {author} {\bibfnamefont {R.}~\bibnamefont {Babbush}}, \emph {et~al.},\ }\bibfield  {title} {\bibinfo {title} {Time-crystalline eigenstate order on a quantum processor},\ }\href {https://doi.org/10.1038/s41586-021-04257-w} {\bibfield  {journal} {\bibinfo  {journal} {Nature}\ }\textbf {\bibinfo {volume} {601}},\ \bibinfo {pages} {531} (\bibinfo {year} {2022})}\BibitemShut {NoStop}%
\bibitem [{\citenamefont {Zhang}\ \emph {et~al.}(2022)\citenamefont {Zhang}, \citenamefont {Jiang}, \citenamefont {Deng}, \citenamefont {Wang}, \citenamefont {Chen}, \citenamefont {Zhang}, \citenamefont {Ren}, \citenamefont {Dong}, \citenamefont {Xu}, \citenamefont {Gao} \emph {et~al.}}]{zhang2022digital}%
  \BibitemOpen
  \bibfield  {author} {\bibinfo {author} {\bibfnamefont {X.}~\bibnamefont {Zhang}}, \bibinfo {author} {\bibfnamefont {W.}~\bibnamefont {Jiang}}, \bibinfo {author} {\bibfnamefont {J.}~\bibnamefont {Deng}}, \bibinfo {author} {\bibfnamefont {K.}~\bibnamefont {Wang}}, \bibinfo {author} {\bibfnamefont {J.}~\bibnamefont {Chen}}, \bibinfo {author} {\bibfnamefont {P.}~\bibnamefont {Zhang}}, \bibinfo {author} {\bibfnamefont {W.}~\bibnamefont {Ren}}, \bibinfo {author} {\bibfnamefont {H.}~\bibnamefont {Dong}}, \bibinfo {author} {\bibfnamefont {S.}~\bibnamefont {Xu}}, \bibinfo {author} {\bibfnamefont {Y.}~\bibnamefont {Gao}}, \emph {et~al.},\ }\bibfield  {title} {\bibinfo {title} {Digital quantum simulation of floquet symmetry-protected topological phases},\ }\href {https://doi.org/10.1038/s41586-022-04854-3} {\bibfield  {journal} {\bibinfo  {journal} {Nature}\ }\textbf {\bibinfo {volume} {607}},\ \bibinfo {pages} {468} (\bibinfo {year} {2022})}\BibitemShut {NoStop}%
\bibitem [{\citenamefont {Jafferis}\ \emph {et~al.}(2022)\citenamefont {Jafferis}, \citenamefont {Zlokapa}, \citenamefont {Lykken}, \citenamefont {Kolchmeyer}, \citenamefont {Davis}, \citenamefont {Lauk}, \citenamefont {Neven},\ and\ \citenamefont {Spiropulu}}]{jafferis2022traversable}%
  \BibitemOpen
  \bibfield  {author} {\bibinfo {author} {\bibfnamefont {D.}~\bibnamefont {Jafferis}}, \bibinfo {author} {\bibfnamefont {A.}~\bibnamefont {Zlokapa}}, \bibinfo {author} {\bibfnamefont {J.~D.}\ \bibnamefont {Lykken}}, \bibinfo {author} {\bibfnamefont {D.~K.}\ \bibnamefont {Kolchmeyer}}, \bibinfo {author} {\bibfnamefont {S.~I.}\ \bibnamefont {Davis}}, \bibinfo {author} {\bibfnamefont {N.}~\bibnamefont {Lauk}}, \bibinfo {author} {\bibfnamefont {H.}~\bibnamefont {Neven}},\ and\ \bibinfo {author} {\bibfnamefont {M.}~\bibnamefont {Spiropulu}},\ }\bibfield  {title} {\bibinfo {title} {Traversable wormhole dynamics on a quantum processor},\ }\href {https://doi.org/10.1038/s41586-022-05424-3} {\bibfield  {journal} {\bibinfo  {journal} {Nature}\ }\textbf {\bibinfo {volume} {612}},\ \bibinfo {pages} {51} (\bibinfo {year} {2022})}\BibitemShut {NoStop}%
\bibitem [{\citenamefont {Ikeda}(2023)}]{ikeda2023first}%
  \BibitemOpen
  \bibfield  {author} {\bibinfo {author} {\bibfnamefont {K.}~\bibnamefont {Ikeda}},\ }\bibfield  {title} {\bibinfo {title} {First realization of quantum energy teleportation on quantum hardware},\ }\href {https://arxiv.org/abs/2301.02666} {\bibfield  {journal} {\bibinfo  {journal} {arXiv:2301.02666}\ } (\bibinfo {year} {2023})}\BibitemShut {NoStop}%
\bibitem [{reg()}]{regime}%
  \BibitemOpen
  \bibinfo {title} {Note that (near) resonant coupling regime represents the coupling strength between elements is much larger than the frequency detuning, while {dispersive} regime means the coupling strength between elements is much smaller than the frequency detuning}\BibitemShut {NoStop}%
\bibitem [{bar()}]{bare}%
  \BibitemOpen
\bibfield  {title} {  }\bibinfo {title} {The bare hamiltonian can be diagonalized to yield the unique normal hamiltonian. in contrast, attempting to derive the bare hamiltonian from the unique inverse transformation of the normal hamiltonian proves unfeasible.}\BibitemShut {Stop}%
\bibitem [{\citenamefont {Nigg}\ \emph {et~al.}(2012)\citenamefont {Nigg}, \citenamefont {Paik}, \citenamefont {Vlastakis}, \citenamefont {Kirchmair}, \citenamefont {Shankar}, \citenamefont {Frunzio}, \citenamefont {Devoret}, \citenamefont {Schoelkopf},\ and\ \citenamefont {Girvin}}]{nigg2012black}%
  \BibitemOpen
\bibfield  {title} {  }\bibfield  {author} {\bibinfo {author} {\bibfnamefont {S.~E.}\ \bibnamefont {Nigg}}, \bibinfo {author} {\bibfnamefont {H.}~\bibnamefont {Paik}}, \bibinfo {author} {\bibfnamefont {B.}~\bibnamefont {Vlastakis}}, \bibinfo {author} {\bibfnamefont {G.}~\bibnamefont {Kirchmair}}, \bibinfo {author} {\bibfnamefont {S.}~\bibnamefont {Shankar}}, \bibinfo {author} {\bibfnamefont {L.}~\bibnamefont {Frunzio}}, \bibinfo {author} {\bibfnamefont {M.}~\bibnamefont {Devoret}}, \bibinfo {author} {\bibfnamefont {R.}~\bibnamefont {Schoelkopf}},\ and\ \bibinfo {author} {\bibfnamefont {S.}~\bibnamefont {Girvin}},\ }\bibfield  {title} {\bibinfo {title} {Black-box superconducting circuit quantization},\ }\href {https://doi.org/https://doi.org/10.1103/PhysRevLett.108.240502} {\bibfield  {journal} {\bibinfo  {journal} {Physical Review Letters}\ }\textbf {\bibinfo {volume} {108}},\ \bibinfo {pages} {240502} (\bibinfo {year} {2012})}\BibitemShut {NoStop}%
\bibitem [{\citenamefont {Minev}\ \emph {et~al.}(2021{\natexlab{a}})\citenamefont {Minev}, \citenamefont {Leghtas}, \citenamefont {Mundhada}, \citenamefont {Christakis}, \citenamefont {Pop},\ and\ \citenamefont {Devoret}}]{minev2021energy}%
  \BibitemOpen
  \bibfield  {author} {\bibinfo {author} {\bibfnamefont {Z.~K.}\ \bibnamefont {Minev}}, \bibinfo {author} {\bibfnamefont {Z.}~\bibnamefont {Leghtas}}, \bibinfo {author} {\bibfnamefont {S.~O.}\ \bibnamefont {Mundhada}}, \bibinfo {author} {\bibfnamefont {L.}~\bibnamefont {Christakis}}, \bibinfo {author} {\bibfnamefont {I.~M.}\ \bibnamefont {Pop}},\ and\ \bibinfo {author} {\bibfnamefont {M.~H.}\ \bibnamefont {Devoret}},\ }\bibfield  {title} {\bibinfo {title} {Energy-participation quantization of josephson circuits},\ }\href {https://doi.org/https://doi.org/10.1038/s41534-021-00461-8} {\bibfield  {journal} {\bibinfo  {journal} {npj Quantum Information}\ }\textbf {\bibinfo {volume} {7}},\ \bibinfo {pages} {1} (\bibinfo {year} {2021}{\natexlab{a}})}\BibitemShut {NoStop}%
\bibitem [{\citenamefont {Paik}\ \emph {et~al.}(2011)\citenamefont {Paik}, \citenamefont {Schuster}, \citenamefont {Bishop}, \citenamefont {Kirchmair}, \citenamefont {Catelani}, \citenamefont {Sears}, \citenamefont {Johnson}, \citenamefont {Reagor}, \citenamefont {Frunzio}, \citenamefont {Glazman} \emph {et~al.}}]{paik2011observation}%
  \BibitemOpen
  \bibfield  {author} {\bibinfo {author} {\bibfnamefont {H.}~\bibnamefont {Paik}}, \bibinfo {author} {\bibfnamefont {D.~I.}\ \bibnamefont {Schuster}}, \bibinfo {author} {\bibfnamefont {L.~S.}\ \bibnamefont {Bishop}}, \bibinfo {author} {\bibfnamefont {G.}~\bibnamefont {Kirchmair}}, \bibinfo {author} {\bibfnamefont {G.}~\bibnamefont {Catelani}}, \bibinfo {author} {\bibfnamefont {A.~P.}\ \bibnamefont {Sears}}, \bibinfo {author} {\bibfnamefont {B.}~\bibnamefont {Johnson}}, \bibinfo {author} {\bibfnamefont {M.}~\bibnamefont {Reagor}}, \bibinfo {author} {\bibfnamefont {L.}~\bibnamefont {Frunzio}}, \bibinfo {author} {\bibfnamefont {L.~I.}\ \bibnamefont {Glazman}}, \emph {et~al.},\ }\bibfield  {title} {\bibinfo {title} {Observation of high coherence in josephson junction qubits measured in a three-dimensional circuit qed architecture},\ }\href {https://doi.org/10.1103/PhysRevLett.107.240501} {\bibfield  {journal} {\bibinfo  {journal} {Physical Review Letters}\ }\textbf {\bibinfo {volume} {107}},\ \bibinfo {pages}
  {240501} (\bibinfo {year} {2011})}\BibitemShut {NoStop}%
\bibitem [{\citenamefont {Krantz}\ \emph {et~al.}(2019)\citenamefont {Krantz}, \citenamefont {Kjaergaard}, \citenamefont {Yan}, \citenamefont {Orlando}, \citenamefont {Gustavsson},\ and\ \citenamefont {Oliver}}]{krantz2019quantum}%
  \BibitemOpen
  \bibfield  {author} {\bibinfo {author} {\bibfnamefont {P.}~\bibnamefont {Krantz}}, \bibinfo {author} {\bibfnamefont {M.}~\bibnamefont {Kjaergaard}}, \bibinfo {author} {\bibfnamefont {F.}~\bibnamefont {Yan}}, \bibinfo {author} {\bibfnamefont {T.~P.}\ \bibnamefont {Orlando}}, \bibinfo {author} {\bibfnamefont {S.}~\bibnamefont {Gustavsson}},\ and\ \bibinfo {author} {\bibfnamefont {W.~D.}\ \bibnamefont {Oliver}},\ }\bibfield  {title} {\bibinfo {title} {A quantum engineer's guide to superconducting qubits},\ }\href {https://doi.org/https://doi.org/10.1063/1.5089550} {\bibfield  {journal} {\bibinfo  {journal} {Applied Physics Reviews}\ }\textbf {\bibinfo {volume} {6}},\ \bibinfo {pages} {021318} (\bibinfo {year} {2019})}\BibitemShut {NoStop}%
\bibitem [{\citenamefont {Blais}\ \emph {et~al.}(2020)\citenamefont {Blais}, \citenamefont {Girvin},\ and\ \citenamefont {Oliver}}]{blais2020quantum}%
  \BibitemOpen
  \bibfield  {author} {\bibinfo {author} {\bibfnamefont {A.}~\bibnamefont {Blais}}, \bibinfo {author} {\bibfnamefont {S.~M.}\ \bibnamefont {Girvin}},\ and\ \bibinfo {author} {\bibfnamefont {W.~D.}\ \bibnamefont {Oliver}},\ }\bibfield  {title} {\bibinfo {title} {Quantum information processing and quantum optics with circuit quantum electrodynamics},\ }\href {https://doi.org/10.1038/s41567-020-0806-z} {\bibfield  {journal} {\bibinfo  {journal} {Nature Physics}\ }\textbf {\bibinfo {volume} {16}},\ \bibinfo {pages} {247} (\bibinfo {year} {2020})}\BibitemShut {NoStop}%
\bibitem [{\citenamefont {Blais}\ \emph {et~al.}(2021)\citenamefont {Blais}, \citenamefont {Grimsmo}, \citenamefont {Girvin},\ and\ \citenamefont {Wallraff}}]{blais2021circuit}%
  \BibitemOpen
  \bibfield  {author} {\bibinfo {author} {\bibfnamefont {A.}~\bibnamefont {Blais}}, \bibinfo {author} {\bibfnamefont {A.~L.}\ \bibnamefont {Grimsmo}}, \bibinfo {author} {\bibfnamefont {S.~M.}\ \bibnamefont {Girvin}},\ and\ \bibinfo {author} {\bibfnamefont {A.}~\bibnamefont {Wallraff}},\ }\bibfield  {title} {\bibinfo {title} {Circuit quantum electrodynamics},\ }\href {https://doi.org/10.1103/RevModPhys.93.025005} {\bibfield  {journal} {\bibinfo  {journal} {Reviews of Modern Physics}\ }\textbf {\bibinfo {volume} {93}},\ \bibinfo {pages} {025005} (\bibinfo {year} {2021})}\BibitemShut {NoStop}%
\bibitem [{\citenamefont {Yan}\ \emph {et~al.}(2018)\citenamefont {Yan}, \citenamefont {Krantz}, \citenamefont {Sung}, \citenamefont {Kjaergaard}, \citenamefont {Campbell}, \citenamefont {Orlando}, \citenamefont {Gustavsson},\ and\ \citenamefont {Oliver}}]{yan2018tunable}%
  \BibitemOpen
  \bibfield  {author} {\bibinfo {author} {\bibfnamefont {F.}~\bibnamefont {Yan}}, \bibinfo {author} {\bibfnamefont {P.}~\bibnamefont {Krantz}}, \bibinfo {author} {\bibfnamefont {Y.}~\bibnamefont {Sung}}, \bibinfo {author} {\bibfnamefont {M.}~\bibnamefont {Kjaergaard}}, \bibinfo {author} {\bibfnamefont {D.~L.}\ \bibnamefont {Campbell}}, \bibinfo {author} {\bibfnamefont {T.~P.}\ \bibnamefont {Orlando}}, \bibinfo {author} {\bibfnamefont {S.}~\bibnamefont {Gustavsson}},\ and\ \bibinfo {author} {\bibfnamefont {W.~D.}\ \bibnamefont {Oliver}},\ }\bibfield  {title} {\bibinfo {title} {Tunable coupling scheme for implementing high-fidelity two-qubit gates},\ }\href {https://doi.org/https://doi.org/10.1103/PhysRevApplied.10.054062} {\bibfield  {journal} {\bibinfo  {journal} {Physical Review Applied}\ }\textbf {\bibinfo {volume} {10}},\ \bibinfo {pages} {054062} (\bibinfo {year} {2018})}\BibitemShut {NoStop}%
\bibitem [{\citenamefont {Yurke}\ and\ \citenamefont {Denker}(1984)}]{yurke1984quantum}%
  \BibitemOpen
  \bibfield  {author} {\bibinfo {author} {\bibfnamefont {B.}~\bibnamefont {Yurke}}\ and\ \bibinfo {author} {\bibfnamefont {J.~S.}\ \bibnamefont {Denker}},\ }\bibfield  {title} {\bibinfo {title} {Quantum network theory},\ }\href {https://doi.org/https://doi.org/10.1103/PhysRevA.29.1419} {\bibfield  {journal} {\bibinfo  {journal} {Physical Review A}\ }\textbf {\bibinfo {volume} {29}},\ \bibinfo {pages} {1419} (\bibinfo {year} {1984})}\BibitemShut {NoStop}%
\bibitem [{\citenamefont {Van~Ruitenbeek}\ \emph {et~al.}(1997)\citenamefont {Van~Ruitenbeek}, \citenamefont {Devoret}, \citenamefont {Esteve},\ and\ \citenamefont {Urbina}}]{van1997conductance}%
  \BibitemOpen
  \bibfield  {author} {\bibinfo {author} {\bibfnamefont {J.}~\bibnamefont {Van~Ruitenbeek}}, \bibinfo {author} {\bibfnamefont {M.}~\bibnamefont {Devoret}}, \bibinfo {author} {\bibfnamefont {D.}~\bibnamefont {Esteve}},\ and\ \bibinfo {author} {\bibfnamefont {C.}~\bibnamefont {Urbina}},\ }\bibfield  {title} {\bibinfo {title} {Conductance quantization in metals: The influence of subband formation on the relative stability of specific contact diameters},\ }\href {https://doi.org/10.1103/PhysRevB.56.12566} {\bibfield  {journal} {\bibinfo  {journal} {Physical Review B}\ }\textbf {\bibinfo {volume} {56}},\ \bibinfo {pages} {12566} (\bibinfo {year} {1997})}\BibitemShut {NoStop}%
\bibitem [{\citenamefont {Burkard}\ \emph {et~al.}(2004)\citenamefont {Burkard}, \citenamefont {Koch},\ and\ \citenamefont {DiVincenzo}}]{burkard2004multilevel}%
  \BibitemOpen
  \bibfield  {author} {\bibinfo {author} {\bibfnamefont {G.}~\bibnamefont {Burkard}}, \bibinfo {author} {\bibfnamefont {R.~H.}\ \bibnamefont {Koch}},\ and\ \bibinfo {author} {\bibfnamefont {D.~P.}\ \bibnamefont {DiVincenzo}},\ }\bibfield  {title} {\bibinfo {title} {Multilevel quantum description of decoherence in superconducting qubits},\ }\href {https://doi.org/https://doi.org/10.1103/PhysRevB.69.064503} {\bibfield  {journal} {\bibinfo  {journal} {Physical Review B}\ }\textbf {\bibinfo {volume} {69}},\ \bibinfo {pages} {064503} (\bibinfo {year} {2004})}\BibitemShut {NoStop}%
\bibitem [{\citenamefont {Malekakhlagh}\ \emph {et~al.}(2017)\citenamefont {Malekakhlagh}, \citenamefont {Petrescu},\ and\ \citenamefont {T{\"u}reci}}]{malekakhlagh2017cutoff}%
  \BibitemOpen
  \bibfield  {author} {\bibinfo {author} {\bibfnamefont {M.}~\bibnamefont {Malekakhlagh}}, \bibinfo {author} {\bibfnamefont {A.}~\bibnamefont {Petrescu}},\ and\ \bibinfo {author} {\bibfnamefont {H.~E.}\ \bibnamefont {T{\"u}reci}},\ }\bibfield  {title} {\bibinfo {title} {Cutoff-free circuit quantum electrodynamics},\ }\href {https://doi.org/https://doi.org/10.1103/PhysRevLett.119.073601} {\bibfield  {journal} {\bibinfo  {journal} {Physical Review Letters}\ }\textbf {\bibinfo {volume} {119}},\ \bibinfo {pages} {073601} (\bibinfo {year} {2017})}\BibitemShut {NoStop}%
\bibitem [{\citenamefont {Gely}\ and\ \citenamefont {Steele}(2020)}]{gely2020qucat}%
  \BibitemOpen
  \bibfield  {author} {\bibinfo {author} {\bibfnamefont {M.~F.}\ \bibnamefont {Gely}}\ and\ \bibinfo {author} {\bibfnamefont {G.~A.}\ \bibnamefont {Steele}},\ }\bibfield  {title} {\bibinfo {title} {Qucat: quantum circuit analyzer tool in python},\ }\href {https://doi.org/10.1088/1367-2630/ab60f6} {\bibfield  {journal} {\bibinfo  {journal} {New Journal of Physics}\ }\textbf {\bibinfo {volume} {22}},\ \bibinfo {pages} {013025} (\bibinfo {year} {2020})}\BibitemShut {NoStop}%
\bibitem [{\citenamefont {Minev}\ \emph {et~al.}(2021{\natexlab{b}})\citenamefont {Minev}, \citenamefont {McConkey}, \citenamefont {Takita}, \citenamefont {Corcoles},\ and\ \citenamefont {Gambetta}}]{minev2021circuit}%
  \BibitemOpen
  \bibfield  {author} {\bibinfo {author} {\bibfnamefont {Z.~K.}\ \bibnamefont {Minev}}, \bibinfo {author} {\bibfnamefont {T.~G.}\ \bibnamefont {McConkey}}, \bibinfo {author} {\bibfnamefont {M.}~\bibnamefont {Takita}}, \bibinfo {author} {\bibfnamefont {A.~D.}\ \bibnamefont {Corcoles}},\ and\ \bibinfo {author} {\bibfnamefont {J.~M.}\ \bibnamefont {Gambetta}},\ }\bibfield  {title} {\bibinfo {title} {Circuit quantum electrodynamics (cqed) with modular quasi-lumped models},\ }\href {https://arxiv.org/abs/2103.10344} {\bibfield  {journal} {\bibinfo  {journal} {arXiv:2103.10344}\ } (\bibinfo {year} {2021}{\natexlab{b}})}\BibitemShut {NoStop}%
\bibitem [{\citenamefont {Bourassa}\ \emph {et~al.}(2012)\citenamefont {Bourassa}, \citenamefont {Beaudoin}, \citenamefont {Gambetta},\ and\ \citenamefont {Blais}}]{bourassa2012josephson}%
  \BibitemOpen
  \bibfield  {author} {\bibinfo {author} {\bibfnamefont {J.}~\bibnamefont {Bourassa}}, \bibinfo {author} {\bibfnamefont {F.}~\bibnamefont {Beaudoin}}, \bibinfo {author} {\bibfnamefont {J.~M.}\ \bibnamefont {Gambetta}},\ and\ \bibinfo {author} {\bibfnamefont {A.}~\bibnamefont {Blais}},\ }\bibfield  {title} {\bibinfo {title} {Josephson-junction-embedded transmission-line resonators: From kerr medium to in-line transmon},\ }\href {https://doi.org/https://doi.org/10.1103/PhysRevA.86.013814} {\bibfield  {journal} {\bibinfo  {journal} {Physical Review A}\ }\textbf {\bibinfo {volume} {86}},\ \bibinfo {pages} {013814} (\bibinfo {year} {2012})}\BibitemShut {NoStop}%
\bibitem [{\citenamefont {Solgun}\ \emph {et~al.}(2014)\citenamefont {Solgun}, \citenamefont {Abraham},\ and\ \citenamefont {DiVincenzo}}]{solgun2014blackbox}%
  \BibitemOpen
  \bibfield  {author} {\bibinfo {author} {\bibfnamefont {F.}~\bibnamefont {Solgun}}, \bibinfo {author} {\bibfnamefont {D.~W.}\ \bibnamefont {Abraham}},\ and\ \bibinfo {author} {\bibfnamefont {D.~P.}\ \bibnamefont {DiVincenzo}},\ }\bibfield  {title} {\bibinfo {title} {Blackbox quantization of superconducting circuits using exact impedance synthesis},\ }\href {https://doi.org/https://doi.org/10.1103/PhysRevB.90.134504} {\bibfield  {journal} {\bibinfo  {journal} {Physical Review B}\ }\textbf {\bibinfo {volume} {90}},\ \bibinfo {pages} {134504} (\bibinfo {year} {2014})}\BibitemShut {NoStop}%
\bibitem [{\citenamefont {Houck}\ \emph {et~al.}(2008)\citenamefont {Houck}, \citenamefont {Schreier}, \citenamefont {Johnson}, \citenamefont {Chow}, \citenamefont {Koch}, \citenamefont {Gambetta}, \citenamefont {Schuster}, \citenamefont {Frunzio}, \citenamefont {Devoret}, \citenamefont {Girvin} \emph {et~al.}}]{houck2008controlling}%
  \BibitemOpen
  \bibfield  {author} {\bibinfo {author} {\bibfnamefont {A.}~\bibnamefont {Houck}}, \bibinfo {author} {\bibfnamefont {J.}~\bibnamefont {Schreier}}, \bibinfo {author} {\bibfnamefont {B.}~\bibnamefont {Johnson}}, \bibinfo {author} {\bibfnamefont {J.}~\bibnamefont {Chow}}, \bibinfo {author} {\bibfnamefont {J.}~\bibnamefont {Koch}}, \bibinfo {author} {\bibfnamefont {J.}~\bibnamefont {Gambetta}}, \bibinfo {author} {\bibfnamefont {D.}~\bibnamefont {Schuster}}, \bibinfo {author} {\bibfnamefont {L.}~\bibnamefont {Frunzio}}, \bibinfo {author} {\bibfnamefont {M.}~\bibnamefont {Devoret}}, \bibinfo {author} {\bibfnamefont {S.}~\bibnamefont {Girvin}}, \emph {et~al.},\ }\bibfield  {title} {\bibinfo {title} {Controlling the spontaneous emission of a superconducting transmon qubit},\ }\href {https://doi.org/10.1103/PhysRevLett.101.080502} {\bibfield  {journal} {\bibinfo  {journal} {Physical Review Letters}\ }\textbf {\bibinfo {volume} {101}},\ \bibinfo {pages} {080502} (\bibinfo {year} {2008})}\BibitemShut {NoStop}%
\bibitem [{\citenamefont {Wang}\ \emph {et~al.}(2016)\citenamefont {Wang}, \citenamefont {Gao}, \citenamefont {Reinhold}, \citenamefont {Heeres}, \citenamefont {Ofek}, \citenamefont {Chou}, \citenamefont {Axline}, \citenamefont {Reagor}, \citenamefont {Blumoff}, \citenamefont {Sliwa} \emph {et~al.}}]{wang2016schrodinger}%
  \BibitemOpen
  \bibfield  {author} {\bibinfo {author} {\bibfnamefont {C.}~\bibnamefont {Wang}}, \bibinfo {author} {\bibfnamefont {Y.~Y.}\ \bibnamefont {Gao}}, \bibinfo {author} {\bibfnamefont {P.}~\bibnamefont {Reinhold}}, \bibinfo {author} {\bibfnamefont {R.~W.}\ \bibnamefont {Heeres}}, \bibinfo {author} {\bibfnamefont {N.}~\bibnamefont {Ofek}}, \bibinfo {author} {\bibfnamefont {K.}~\bibnamefont {Chou}}, \bibinfo {author} {\bibfnamefont {C.}~\bibnamefont {Axline}}, \bibinfo {author} {\bibfnamefont {M.}~\bibnamefont {Reagor}}, \bibinfo {author} {\bibfnamefont {J.}~\bibnamefont {Blumoff}}, \bibinfo {author} {\bibfnamefont {K.}~\bibnamefont {Sliwa}}, \emph {et~al.},\ }\bibfield  {title} {\bibinfo {title} {A schr{\"o}dinger cat living in two boxes},\ }\href {https://doi.org/10.1126/science.aaf2941} {\bibfield  {journal} {\bibinfo  {journal} {Science}\ }\textbf {\bibinfo {volume} {352}},\ \bibinfo {pages} {1087} (\bibinfo {year} {2016})}\BibitemShut {NoStop}%
\bibitem [{\citenamefont {Teoh}\ \emph {et~al.}(2023)\citenamefont {Teoh}, \citenamefont {Winkel}, \citenamefont {Babla}, \citenamefont {Chapman}, \citenamefont {Claes}, \citenamefont {de~Graaf}, \citenamefont {Garmon}, \citenamefont {Kalfus}, \citenamefont {Lu}, \citenamefont {Maiti} \emph {et~al.}}]{teoh2023dual}%
  \BibitemOpen
  \bibfield  {author} {\bibinfo {author} {\bibfnamefont {J.~D.}\ \bibnamefont {Teoh}}, \bibinfo {author} {\bibfnamefont {P.}~\bibnamefont {Winkel}}, \bibinfo {author} {\bibfnamefont {H.~K.}\ \bibnamefont {Babla}}, \bibinfo {author} {\bibfnamefont {B.~J.}\ \bibnamefont {Chapman}}, \bibinfo {author} {\bibfnamefont {J.}~\bibnamefont {Claes}}, \bibinfo {author} {\bibfnamefont {S.~J.}\ \bibnamefont {de~Graaf}}, \bibinfo {author} {\bibfnamefont {J.~W.}\ \bibnamefont {Garmon}}, \bibinfo {author} {\bibfnamefont {W.~D.}\ \bibnamefont {Kalfus}}, \bibinfo {author} {\bibfnamefont {Y.}~\bibnamefont {Lu}}, \bibinfo {author} {\bibfnamefont {A.}~\bibnamefont {Maiti}}, \emph {et~al.},\ }\bibfield  {title} {\bibinfo {title} {Dual-rail encoding with superconducting cavities},\ }\href {https://doi.org/10.1073/pnas.2221736120} {\bibfield  {journal} {\bibinfo  {journal} {Proceedings of the National Academy of Sciences}\ }\textbf {\bibinfo {volume} {120}},\ \bibinfo {pages} {e2221736120} (\bibinfo {year} {2023})}\BibitemShut
  {NoStop}%
\bibitem [{\citenamefont {Dubyna}\ and\ \citenamefont {Kuo}(2020)}]{dubyna2020inter}%
  \BibitemOpen
  \bibfield  {author} {\bibinfo {author} {\bibfnamefont {D.}~\bibnamefont {Dubyna}}\ and\ \bibinfo {author} {\bibfnamefont {W.}~\bibnamefont {Kuo}},\ }\bibfield  {title} {\bibinfo {title} {Inter-qubit interaction mediated by collective modes in a linear array of three-dimensional cavities},\ }\href {https://doi.org/10.1088/2058-9565/ab8114} {\bibfield  {journal} {\bibinfo  {journal} {Quantum Science and Technology}\ }\textbf {\bibinfo {volume} {5}},\ \bibinfo {pages} {035002} (\bibinfo {year} {2020})}\BibitemShut {NoStop}%
\bibitem [{\citenamefont {Hornibrook}\ \emph {et~al.}(2012)\citenamefont {Hornibrook}, \citenamefont {Mitchell},\ and\ \citenamefont {Reilly}}]{hornibrook2012superconducting}%
  \BibitemOpen
  \bibfield  {author} {\bibinfo {author} {\bibfnamefont {J.}~\bibnamefont {Hornibrook}}, \bibinfo {author} {\bibfnamefont {E.}~\bibnamefont {Mitchell}},\ and\ \bibinfo {author} {\bibfnamefont {D.}~\bibnamefont {Reilly}},\ }\bibfield  {title} {\bibinfo {title} {Superconducting resonators with parasitic electromagnetic environments},\ }\href {https://arxiv.org/abs/1203.4442} {\bibfield  {journal} {\bibinfo  {journal} {arXiv:1203.4442}\ } (\bibinfo {year} {2012})}\BibitemShut {NoStop}%
\bibitem [{\citenamefont {Mundada}\ \emph {et~al.}(2019)\citenamefont {Mundada}, \citenamefont {Zhang}, \citenamefont {Hazard},\ and\ \citenamefont {Houck}}]{mundada2019suppression}%
  \BibitemOpen
  \bibfield  {author} {\bibinfo {author} {\bibfnamefont {P.}~\bibnamefont {Mundada}}, \bibinfo {author} {\bibfnamefont {G.}~\bibnamefont {Zhang}}, \bibinfo {author} {\bibfnamefont {T.}~\bibnamefont {Hazard}},\ and\ \bibinfo {author} {\bibfnamefont {A.}~\bibnamefont {Houck}},\ }\bibfield  {title} {\bibinfo {title} {Suppression of qubit crosstalk in a tunable coupling superconducting circuit},\ }\href {https://doi.org/10.1103/PhysRevApplied.12.054023} {\bibfield  {journal} {\bibinfo  {journal} {Physical Review Applied}\ }\textbf {\bibinfo {volume} {12}},\ \bibinfo {pages} {054023} (\bibinfo {year} {2019})}\BibitemShut {NoStop}%
\bibitem [{\citenamefont {Ku}\ \emph {et~al.}(2020)\citenamefont {Ku}, \citenamefont {Xu}, \citenamefont {Brink}, \citenamefont {McKay}, \citenamefont {Hertzberg}, \citenamefont {Ansari},\ and\ \citenamefont {Plourde}}]{ku2020suppression}%
  \BibitemOpen
  \bibfield  {author} {\bibinfo {author} {\bibfnamefont {J.}~\bibnamefont {Ku}}, \bibinfo {author} {\bibfnamefont {X.}~\bibnamefont {Xu}}, \bibinfo {author} {\bibfnamefont {M.}~\bibnamefont {Brink}}, \bibinfo {author} {\bibfnamefont {D.~C.}\ \bibnamefont {McKay}}, \bibinfo {author} {\bibfnamefont {J.~B.}\ \bibnamefont {Hertzberg}}, \bibinfo {author} {\bibfnamefont {M.~H.}\ \bibnamefont {Ansari}},\ and\ \bibinfo {author} {\bibfnamefont {B.}~\bibnamefont {Plourde}},\ }\bibfield  {title} {\bibinfo {title} {Suppression of unwanted zz interactions in a hybrid two-qubit system},\ }\href {https://doi.org/10.1103/PhysRevLett.125.200504} {\bibfield  {journal} {\bibinfo  {journal} {Physical Review Letters}\ }\textbf {\bibinfo {volume} {125}},\ \bibinfo {pages} {200504} (\bibinfo {year} {2020})}\BibitemShut {NoStop}%
\bibitem [{\citenamefont {Sung}\ \emph {et~al.}(2021)\citenamefont {Sung}, \citenamefont {Ding}, \citenamefont {Braum\"uller}, \citenamefont {Veps\"al\"ainen}, \citenamefont {Kannan}, \citenamefont {Kjaergaard}, \citenamefont {Greene}, \citenamefont {Samach}, \citenamefont {McNally}, \citenamefont {Kim} \emph {et~al.}}]{sung2021realization}%
  \BibitemOpen
  \bibfield  {author} {\bibinfo {author} {\bibfnamefont {Y.}~\bibnamefont {Sung}}, \bibinfo {author} {\bibfnamefont {L.}~\bibnamefont {Ding}}, \bibinfo {author} {\bibfnamefont {J.}~\bibnamefont {Braum\"uller}}, \bibinfo {author} {\bibfnamefont {A.}~\bibnamefont {Veps\"al\"ainen}}, \bibinfo {author} {\bibfnamefont {B.}~\bibnamefont {Kannan}}, \bibinfo {author} {\bibfnamefont {M.}~\bibnamefont {Kjaergaard}}, \bibinfo {author} {\bibfnamefont {A.}~\bibnamefont {Greene}}, \bibinfo {author} {\bibfnamefont {G.~O.}\ \bibnamefont {Samach}}, \bibinfo {author} {\bibfnamefont {C.}~\bibnamefont {McNally}}, \bibinfo {author} {\bibfnamefont {D.}~\bibnamefont {Kim}}, \emph {et~al.},\ }\bibfield  {title} {\bibinfo {title} {Realization of high-fidelity cz and $zz$-free iswap gates with a tunable coupler},\ }\href {https://doi.org/10.1103/PhysRevX.11.021058} {\bibfield  {journal} {\bibinfo  {journal} {Phys. Rev. X}\ }\textbf {\bibinfo {volume} {11}},\ \bibinfo {pages} {021058} (\bibinfo {year} {2021})}\BibitemShut {NoStop}%
\bibitem [{\citenamefont {Wenner}\ \emph {et~al.}(2011)\citenamefont {Wenner}, \citenamefont {Barends}, \citenamefont {Bialczak}, \citenamefont {Chen}, \citenamefont {Kelly}, \citenamefont {Lucero}, \citenamefont {Mariantoni}, \citenamefont {Megrant}, \citenamefont {O’Malley}, \citenamefont {Sank} \emph {et~al.}}]{wenner2011surface}%
  \BibitemOpen
  \bibfield  {author} {\bibinfo {author} {\bibfnamefont {J.}~\bibnamefont {Wenner}}, \bibinfo {author} {\bibfnamefont {R.}~\bibnamefont {Barends}}, \bibinfo {author} {\bibfnamefont {R.}~\bibnamefont {Bialczak}}, \bibinfo {author} {\bibfnamefont {Y.}~\bibnamefont {Chen}}, \bibinfo {author} {\bibfnamefont {J.}~\bibnamefont {Kelly}}, \bibinfo {author} {\bibfnamefont {E.}~\bibnamefont {Lucero}}, \bibinfo {author} {\bibfnamefont {M.}~\bibnamefont {Mariantoni}}, \bibinfo {author} {\bibfnamefont {A.}~\bibnamefont {Megrant}}, \bibinfo {author} {\bibfnamefont {P.}~\bibnamefont {O’Malley}}, \bibinfo {author} {\bibfnamefont {D.}~\bibnamefont {Sank}}, \emph {et~al.},\ }\bibfield  {title} {\bibinfo {title} {Surface loss simulations of superconducting coplanar waveguide resonators},\ }\href {https://doi.org/10.1063/1.3637047} {\bibfield  {journal} {\bibinfo  {journal} {Applied Physics Letters}\ }\textbf {\bibinfo {volume} {99}},\ \bibinfo {pages} {113513} (\bibinfo {year} {2011})}\BibitemShut {NoStop}%
\bibitem [{\citenamefont {Minev}\ \emph {et~al.}(2021{\natexlab{c}})\citenamefont {Minev}, \citenamefont {Leghtas}, \citenamefont {Reinhold}, \citenamefont {Mundhada}, \citenamefont {Diringer}, \citenamefont {Hillel}, \citenamefont {Wang}, \citenamefont {Facchini}, \citenamefont {Shah},\ and\ \citenamefont {Devoret}}]{pyEPR}%
  \BibitemOpen
  \bibfield  {author} {\bibinfo {author} {\bibfnamefont {Z.~K.}\ \bibnamefont {Minev}}, \bibinfo {author} {\bibfnamefont {Z.}~\bibnamefont {Leghtas}}, \bibinfo {author} {\bibfnamefont {P.}~\bibnamefont {Reinhold}}, \bibinfo {author} {\bibfnamefont {S.~O.}\ \bibnamefont {Mundhada}}, \bibinfo {author} {\bibfnamefont {A.}~\bibnamefont {Diringer}}, \bibinfo {author} {\bibfnamefont {D.~C.}\ \bibnamefont {Hillel}}, \bibinfo {author} {\bibfnamefont {D.~Z.-R.}\ \bibnamefont {Wang}}, \bibinfo {author} {\bibfnamefont {M.}~\bibnamefont {Facchini}}, \bibinfo {author} {\bibfnamefont {P.~A.}\ \bibnamefont {Shah}},\ and\ \bibinfo {author} {\bibfnamefont {M.}~\bibnamefont {Devoret}},\ }\href {https://doi.org/10.5281/zenodo.4744447} {\bibinfo {title} {{pyEPR: The energy-participation-ratio (EPR) open- source framework for quantum device design}}} (\bibinfo {year} {2021}{\natexlab{c}})\BibitemShut {NoStop}%
\bibitem [{\citenamefont {Yanay}\ \emph {et~al.}(2022)\citenamefont {Yanay}, \citenamefont {Braum{\"u}ller}, \citenamefont {Orlando}, \citenamefont {Gustavsson}, \citenamefont {Tahan},\ and\ \citenamefont {Oliver}}]{yanay2022mediated}%
  \BibitemOpen
  \bibfield  {author} {\bibinfo {author} {\bibfnamefont {Y.}~\bibnamefont {Yanay}}, \bibinfo {author} {\bibfnamefont {J.}~\bibnamefont {Braum{\"u}ller}}, \bibinfo {author} {\bibfnamefont {T.~P.}\ \bibnamefont {Orlando}}, \bibinfo {author} {\bibfnamefont {S.}~\bibnamefont {Gustavsson}}, \bibinfo {author} {\bibfnamefont {C.}~\bibnamefont {Tahan}},\ and\ \bibinfo {author} {\bibfnamefont {W.~D.}\ \bibnamefont {Oliver}},\ }\bibfield  {title} {\bibinfo {title} {Mediated interactions beyond the nearest neighbor in an array of superconducting qubits},\ }\href {https://doi.org/10.1103/PhysRevApplied.17.034060} {\bibfield  {journal} {\bibinfo  {journal} {Physical Review Applied}\ }\textbf {\bibinfo {volume} {17}},\ \bibinfo {pages} {034060} (\bibinfo {year} {2022})}\BibitemShut {NoStop}%
\end{thebibliography}%

\end{document}